\documentstyle[12pt, epsf]{article}
\begin{document}

\begin{center}
{\large \bf
Asymmetric shape and dynamic stability 
} 
\end{center}
\begin{center}
{\large \bf 
of exciton-phonon solitons moving in a periodic medium
} 
\end{center}
\vspace*{0.3cm}
\begin{center}
Dan Roubtsov\,$^{\dag}$ \,and\, Yves L\'epine
\end{center}
\begin{center}
{\it 
Groupe de Recherche en Physique et Technologie des Couches Minces, 
}
\end{center}
\begin{center}
{\it 
D\'epartement de Physique, Universit\'e de Montr\'eal, 
}
\end{center}
\begin{center}
C.P. 6128, succ. Centre-ville, Montr\'eal, QC, H3C\,3J7, Canada 
\end{center}
\vspace*{0.2cm}
\begin{center}
E. Nihan {\"O}nder
\end{center}
\begin{center}
{\it 
Institut de g\'enie nucl\'eaire, {\'E}cole 
 Polytechnique de Montr\'eal,
}
\end{center}
\begin{center}
C.P. 6079, succ. Centre-ville, Montr\'eal, QC, H3C\,3A7, Canada 
\end{center}
\begin{center}
$^{\dag}$\,e-mail: roubtsod@magellan.umontreal.ca 
\end{center}
\vspace*{0.4cm}

\begin{abstract}

Solitons are known to move ballistically through a medium 
without changement of their shape.
In  practice, the shape of moving inhomogeneous states changes,  
and a long lasting tail appears 
behind the soliton moving in a periodic medium.
Such a behavior can be described  
within the model of exciton-phonon coherent states   
as a dynamic effect in soliton transport.  
We argue that  
the coupling   
between  bosonic excitations of the medium, such as excitons,
and elastic modes of it, such as phonons,  
can be responsible for these effects. 
We derive nonlinear dynamic equations for the excited medium 
in the long wavelength approximation (a generalized Zakharov system) 
and apply  
a kind of ballistic ansatz for the
coherent state of bosons and displacement field. 
%
%
Like solitons, the quasistationary solution 
we obtain on this way can move
through the medium ballistically. Unlike solitons and kinks, there are  
inhomogeneous corrections to the 
ballistic velocity and the coherent phase of the condensate that  
control the changement of the packet shape.
In the limit of $T\rightarrow 0$, 
laws 
of conservation prescribe formation of the cloud of 
collective excitations around the 
quasistationary Bose-core of the packet, and 
the density of a growing boson-phonon tail  
behind the
moving coherent part can be estimated. 
The total packet can be associated with an {\it exciton-phonon comet}
with the quasistable coherent Bose-core and incoherent tail
moving in the medium. 
\end{abstract}
\vspace*{0.5cm}


\vspace*{0.5cm}
submitted to Phys. Rev. B, \,\,\,preprint at \,cond-mat/0008284,
\,\,\,revised October 2000

\newpage

\section{Introduction}

Some localized solutions of a special (and, actually, quite a narrow) 
class of nonlinear evolution equations 
are known to be able to move
 conserving their
shape and interacting with each other like  particles. 
The typical examples are the solitons of Nonlinear Schr\"odinger equation (NLS) and Korteveg-de-Vriese equation 
(KdV), 
the kinks of sine-Gordon equation (SG) to name a few, 
\cite{Dodd},\cite{Infeld}. It is worthy of mentioning that
these particular solutions conserve all the intergals of motions prescribed by 
dynamic equations.
However, as the evolution 
equations mentioned above are the result of approximation 
of more complicated dynamic equations, 
the solitonic properties of their solutions are the approximation as well.
The question up to which extent they survive in `reality' leads to 
interesting physical problems  
that urge people to go beyond the theory of exactly solvable models 
\cite{Bishop},\cite{Enns},\cite{Pelin}.

Fortunately, the exact results obtained within   
refine mathematical models can be compared with the real physical experiments. 
In addition, numerical simulations within more realistic models are performed
for many equations supporting solitonic solutions  
\cite{numeric}.
As a result, 
one can realize under which conditions the concept of soliton 
is a good approximation
to interpret physical data.   
For example, the NLS equation is used  to describe the phenomenon of 
Bose-Einstein condensation of dilute 
trapped gases
\cite{review} and sharp pulses of light in optical fibers \cite{Newell},  
and the KdV equation is used in hydrodynamics 
to model the moving packets of surface waves 
in channels \cite{Benjamin}. The SG equation models surface grow and 
reconstruction in material science \cite{Braun}. 

In this article, we discuss the transport properties of  excited states 
(generally of electron transition nature) 
in a  medium (a periodic structure, such as a crystal or 
semiconductor structure, some periodic biological tissues, etc.). 
Under certain conditions, a coherent state, which involves  
both the electronic (or spin) excitations and the medium elastic excitations,  
can appear in such structures. This state turns out to be localized 
inside the medium  
and can travel ballistically through it. 
In Biological Physics, such a state is known as 
 Davydov soliton \cite{Davydov}, whereas in Condensed Matter Physics, 
one can mention, for example, 
Spin-Peierls Systems \cite{Spin}, Charge Density Waves \cite{CDW}, 
and anomalous transport of excitonic packets in semiconducting crystals and 
heterostructures \cite{excitons}. In particular, we are motivated by 
the ballistic transport of excitons 
in pure 3D crystals, such as \, Cu$_{2}$0 \cite{cuprous}. 
However,  
the ballistic transport of localized coherent structures
and a set of nonlinear equations being used to describe it 
can be found in modeling of different physical phenomena \cite{Infeld},
\cite{in_action}, \cite{Tusin}. For example, modeling of the coherent excited states of 
electronic origin ends up
with NLS equation, whereas the coherent states of the elastic medium are 
described by 
the wave equation with nonlinear anharmonic terms (NLW equation) 
\cite{CMP},\cite{Loutsenko},\cite{Brasil}.
In plasma physics, this is the case of Zakharov system of equations 
\cite{Zakharov},\cite{Sulem},\cite{Berge}, while 
the Davey-Stewartson system can be mentioned in the context of 
fluid mechanics \cite{NMPZ}.
Note that the two equations comprising such a system, NLS and NLW,  are quite different 
from the mathematical point of view, but  
they are coupled with each other \cite{NMPZ},\cite{Ablowitz},\cite{Melnik} and 
this is the origin of the wealth of physical effects 
they can describe.   

Not surprisingly for the moving soliton-like states in periodic media,
a shape of experimentally registered signals is far from being look like 
a ``true'' soliton (e.g., \,$\propto 1/{\rm cosh}\bigl(\,L_{0}^{-1}(x-vt)\,\bigr)$\,) 
or  kink. One can assume that  
nonzero temperature,  scattering on impurities
and just a noise factor can be responsible for this. 
In some cases, however, the character of changement hints on some
regular dynamic reason for these effects. 
For example, 
an asymmetric form of the signal with the pronounced sharp front and 
a long lasting tail  behind the single soliton are observed on experiment 
\cite{cuprous},\cite{tail_exp},\cite{bio}. 
Note that the subsonic ballistic transport is under consideration, i.e., \,
$\langle v \rangle < c_{s}$, where $\langle v \rangle $ 
is the speed of the packet and $c_{s}$ is the (longitudinal) sound
speed in the periodic medium. 
The similar effects, such as delocalized solitons,  were also observed in 
different physical settings. For example, in a kind of pump and probe experiments 
with sharp pulses of light in fibers, the solitonic shape of a probe pulse was significantly changed 
by Stimulated Raman Scattering \cite{Agrawal}; see also \cite{Kaup}. 
In fluid mechanics, one can mention the problem of wave breaking and 
appearance of the so-called 
peakons and compactons in the soliton-like solutions of 
generalized KdV equations \cite{Camassa}; see also \cite{Boyd}. 
In essence, the non-steady solitons    
are found to be able to move with an
acceleration and generate ``tails'' behind themselves.

In contrast, the transition to diffusion regime is possible
for the ballistic exciton-phonon droplets, 
and, in the case of amplification of the soliton-like state formed
inside the exciton-phonon packet, 
its ballistic velocity is found to be almost unchanged
during such a non-stationary process \cite{cuprous}.
These facts hint at different physics controlling the 
moving coherent state
formed by excitons and phonons.

In this article, we assume that the coupling between collective
excitations of the periodic medium can be responsible for such 
effects
in Condensed 
Matter Physics, and 
a proper dynamic model 
can reproduce them even at $T\rightarrow 0$. 
To support this hypothesis, we consider a field model that admits
a soliton-like solution 
and can be easily generalized without a loss of physical clarity. 

This is a two field model, in which the excitations of the medium are modeled 
by a nonideal
Bose-gas coupled with long wavelength modes of the displacement field.
For example, two (or even several) interacting boson fields can model 
different phenomena  
in Condensed Matter Physics \cite{CMP}.
In the case of semiconductors, one can often disregard the influence of  
free fermions 
or fermionic complexes 
on the processes under consideration.  
It turns out that many processes involving photons, excitons, and phonons   
can be  described by use of the  language of interacting Bose-fields \cite{Inoue}.
Moreover, if there is a branch of optically inactive excitons in a crystal, one 
can even exclude the 
photons from simple models dealing with such excitons. 
In addition, the lifetime of a  moving exciton with $\hbar k_{0} \simeq m_{\rm x}c_{s}$ 
can be large enough  
in semiconductor structures,
so that the transport properties of a packet of moving excitons  
can be observed experimentally \cite{cuprous},\cite{exc_transp}.

\section{Hamiltonian of the model}

The Hamiltonian of the medium can be taken in the following form:
\begin{equation}
\hat{H}=H_{\rm gas}( \hat{\psi},\,\hat{\psi}^{\dag})+ 
   {H_{\rm ph} }(\hat{u},\,\hat{\pi})+ 
   {H_{\rm int} }(\hat{u},\,\hat{\psi}^{\dag} \hat{\psi}).
\label{ham}
\end{equation}
Here, $\hat{\psi}$, $\hat{\psi}^{\dag}$ are  the Bose-gas field operators; they  
stands in the non-relativistic Hamiltonian $H_{\rm gas}$, 
whereas 
$\hat{u}$ is the 
displacement field operator and  $\hat{\pi}$ is the momentum density 
operator
conjugate to $\hat{u}$. This pair 
stands  in a Phonon Hamiltonian 
${H_{\rm ph} }$.  We consider  
a nonideal gas of phonons in the long-wavelength approximation and take
into  account the acoustic branch only, 
e.g., the medium is a  3D crystal of  the volume $V= L\,S_{\perp}$.

The Bose-gas Hamiltonian has the following
form:
$$
H_{\rm gas}
=\,\!\int\!d{\bf x} \, \tilde{E}_{g}\, \hat{\psi}^{\dag}\hat{\psi}({\bf x}) + 
\frac{\hbar^{2}}{2m}\nabla \hat{\psi}^{\dag}\,\nabla\hat{\psi}({\bf x}) +
\frac{ \nu_{0} }{2}\, \Bigl(\hat{\psi}^{\dag}({\bf x})\Bigr)^{2} 
\Bigl(\hat{\psi}({\bf x})\Bigr)^{2}
 +
\frac{ \nu_{1} }{3}\,
\Bigl(\hat{\psi}^{\dag}({\bf x})\Bigr)^{3} 
\Bigl(\hat{\psi}({\bf x})\Bigr)^{3},
$$
where $\nu_{0} > 0$ is the strength of  two particle repulsive interaction, 
and $\nu_{1}>0$
is the strength of three particle one.   
As the two particle interaction ($\sim \nu_{0} \,(\psi^{\dag} )^{2} \psi^{2} $\,)
can be strongly renormalized because of 
interaction with other fields, 
we include  
the  hard core interaction term (modeled by repulsion
\,$\sim \nu_{1}\,(\psi^{\dag} )^{3} \psi^{3}$  in $H_{\rm gas}$).
We assume  that the `bare'
characteristic energies of the  particle-particle interactions satisfy 
the following inequality:
$$
0< \nu_{1}/{a}_{\rm x}^{6} 
< (\ll)\,
\nu_{0}/{a}_{\rm x}^{3} \simeq  {\rm const}\,{\rm Ry}_{\rm x},
\,\,\,\,\,\,{\rm const} \simeq 10,
$$  
see \cite{ScLength} for discussion.
Here, ${a}_{\rm x}$ and ${\rm Ry}_{\rm x}$ are the exciton Bohr radius and characteristic Rydberg
energy, respectively. We count  the energy of a  free particle from \,$
\tilde{E}_{g}= {E}_{g}-{\rm Ry}_{\rm x}>0$, \,so that \,$E_{\bf k} =
\tilde{E}_{g} + \hbar {\bf k}^{2}/2m$. (For semiconducting materials, ${E}_{g}$ is the
fundamental gap.)
    
As only the longitudinal phonons  ${\bf u}_{l}$ 
interact with the Bose-field $\hat{\psi}$, $\hat{\psi}^{\dag}$ in our model,
one can exclude the transversal phonons ($\nabla {\bf u}_{t} = 0$)
from $H_{\rm ph}$.
Then, it can be  reduced to the following simple form:
$$
H_{\rm ph}=\int\! \frac{ \hat{
\pi }^{2}({\bf x})}{2\rho 
} \,+ \,\frac{\rho c_{l}^{2} }{2}\,
\partial_{j}\hat{u}_{s}
\partial_{j}
\hat{u}_{s}({\bf x}) +
\frac {\rho c_{l}^{2}  }{3}\,\kappa_{3}\,
\partial_{ j}\hat{u}_{s}
\partial_{j}
\hat{u}_{s}
\partial_{j}
\hat{u}_{s}({\bf x}) + 
\frac {\rho c_{l}^{2}  }{4}\,\kappa_{4}\,
\bigl(\partial_{ j}\hat{u}_{s}\bigr)^{4}
\,d{\bf x}, 
$$
where $c_{l}$ is the longitudinal sound speed of the crystal, and
the  dimensionless parameters $\kappa_{3}$ and $\kappa_{4}$ account
for cubic and quartic nonlinearities.
(In this article, we will not take into account a quartic
term, so \,$\kappa_{4}\simeq 0$.)

We take the gas-phonon interaction in the 
form of Deformation Potential:
\begin{equation}
 {H}_{\rm int}=\int \!\sigma_{0}\, 
  \partial_{j}\hat{ u}_{j}({\bf x})\, \hat{\psi}^{\dag}\hat{\psi} 
({\bf x})  + \vartheta_{0}\,\partial_{j}\hat{ u}_{j}({\bf x})\, 
\Bigl(-\hat{\psi}^{\dag} \Delta\,\hat{\psi}({\bf x})\,\Bigr) \,d{\bf x}, 
\label{twoT}
\end{equation} 
where $\sigma_{0}$  and $\vartheta_{0}$   are the coupling constants.
Note that this Hamiltonian is equivalent to 
\begin{equation}
\tilde{H}_{\rm int}=\int \!\sigma_{0}\, 
  \partial_{j}\hat{ u}_{j}({\bf x})\, \hat{\psi}^{\dag}\hat{\psi} 
({\bf x})  + \vartheta_{0}\, \partial_{j}\hat{ u}_{j}({\bf x})\, 
\nabla \hat{\psi}^{\dag}\, \nabla \hat{\psi}({\bf x}) \,d{\bf x}.
\label{twoT1}
\end{equation}         
Here, we choose \,$\sigma_{0}>0$ \,and do not fix the sign of $\vartheta_{0}$.
Developing a theory, we have a freedom with the sign  of $\vartheta_{0}$; 
its value, however,  can be roughly estimated as 
$|\vartheta_{0}| \simeq \hbar^{2}/2m$. 
(Recall that $\sigma_{0} \sim E_{g}$.) 

If the exciton Bohr radius ${a}_{\rm x}$ is no more than several times larger than 
the lattice constant
$a_{l}$, the second term in (\ref{twoT}) can be important too.
(In fact, this is an intermediate case between  
Frenkel exciton and Wannier one with ${a}_{\rm x}\gg a_{l}$.)
Then, the following coupling terms appear in Heisenberg equations
\begin{equation}
 \tilde{E}_{g} \hat{\psi} \, \rightarrow \,  \tilde{E}_{g} \hat{\psi}  \,+\, 
\sigma_{0}\,\partial_{j}\hat{ u}_{j}\,\hat{\psi}, 
\end{equation}
\begin{equation}
 -(\hbar^{2}/2m)\,\Delta \hat{\psi}  \, \rightarrow \,
 - (\hbar^{2}/2m)\,\Delta \hat{\psi} \,- 
 \, \vartheta_{0}\,\partial_{j}\hat{ u}_{j}\,\Delta \hat{\psi}. 
\end{equation}
In terms of the lattice analog of the medium Hamiltonian, 
we made the hopping matrix element $t$
\,($t_{ij}$ in \,$\hat{H}_{\rm gas}  
\sim t_{ij}\,\hat{\psi}_{i}^{\dag}\hat{\psi}_{j}$)
of the `lattice' boson $\hat{\psi}_{j}$ to be 
dependent on the
`lattice' deformation field $\hat{u}_{j}$, 
$$
t_{ij}\rightarrow t_{ij}(u_{i},\,u_{j})\approx t_{ij}+ 
\tilde{\vartheta}_{0}\,(u_{i}-u_{j}).
$$
Meanwhile, the energy on a cite, $\varepsilon_{0,\,i}=
\varepsilon_{0}$ (in 
$\hat{H}_{\rm gas} 
\sim \varepsilon_{0,\,i}\,\hat{\psi}_{i}^{\dag}\hat{\psi}_{i}$)\,
depends on the lattice displacements too, f.ex.,
$$ 
\varepsilon_{0,\,i}\rightarrow \varepsilon_{0} + 
\tilde{\sigma}_{0}\,(u_{i+1}-u_{i-1}), 
$$
see \cite{Hennig} for discussion.

\section{Dynamic equations}

Our aim is to investigate a special class of solutions of  
Hamiltonian (\ref{ham}), namely, 
we will search for the localized moving excitations (or packets).
In the case of $\vartheta_{0}=0$, such a solution exists and  
the packet can move through the crystal 
with the constant velocity $v$
saving its shape in a manner as the solitons and kinks do in nonlinear media 
\cite{Davydov},\cite{Loutsenko}.
To simplify the equations of motion, we choose the quasi-1D approximation.
This means  
$\hat{\psi}({\bf x},t) \rightarrow \hat{\psi}({ x},t)$, 
$\hat{ \bf u}({\bf x},t) \rightarrow \hat{ u}_{x}({ x},t)$.
In fact,  we assume that the packet is  inhomogeneous along the  
$Ox$ axis only, and \,$v \parallel Ox$. 
Thus,  it occupies all the area of \,$S_{\perp}$.  
(For a discussion on the validity of 1D approximation, see \cite{RL}.) 
Then, we can write the Heisenberg equations of motion as 
$$
i\hbar\,\partial_{t}\hat{\psi}({ x},t)\,= 
$$
\begin{equation}
=\,\Bigr(\tilde{E}_{g} - {{\hbar^{2}}\over{2m}}\partial_{x}^{2} 
- \vartheta_{0}\,\partial_{x}\hat{ u}_{x} ({ x},t)\,\partial_{x}^{2} 
+ \nu_{0}
\hat{\psi}^{\dag}\hat{\psi}({ x},t)+ 
\nu_{1}\hat{\psi}^{\dag \,2}\hat{\psi}^{2}({ x},t)\,\Bigl)  
\hat{\psi}({ x},t) + \sigma_{0}\,\partial_{j} \hat{ u}_{j} ({ x},t)\,
\hat{\psi}({ x},t),
\label{eq1M}
\end{equation}
$$
\Bigl( \,\partial_{t}^{2} -  
c_{l}^{2}\partial_{x}^{2} \,\Bigr) \hat{ u}_{x}({ x},t)\,-
$$ 
\begin{equation}
-\,c_{l}^{2}
2 \kappa_{3}\,\partial_{x}^{2} \hat{ u}_{x}\,\partial_{x}\hat{ u}_{x}({ x},t) 
\,=
\rho^{-1}\sigma_{0}\,\partial_{x}\bigl(\hat{\psi}^{\dag}\hat{\psi}({ x},t)\bigr)
+\rho^{-1}\vartheta_{0}\,\partial_{x}\Bigl(\partial_{x}\hat{\psi}^{\dag}\,\partial_{x}\hat{\psi}\Bigr).
\label{eq2M}
\end{equation}

After some energy (and momentum) was pumped into the medium, 
an excited state of it appears near the boundary 
where the external energy (and momentum) was absorbed, see Fig. 1. 
Due to initial conditions, 
the excited state is a localized one, and it can be   
modeled as a droplet consisting of  
excitons (Bose-particles) and  phonons. Moreover, it 
can acquire an average momentum directed along the 
$Ox$ axis because of unidirectional phonon production 
during the thermalization stage.   
Thus, the exciton-phonon droplet starts to move \cite{STikho}.   
The important assumption, however, is the appearance of a  moving {\it coherent} field, 
or a condensate, from the localized excited state of a medium.
For example, if the medium is a 3D crystal or 
an array of channels embedded into a matrix,  
it has to be coupled with a thermostat at low temperature. 
Moreover, the medium has to posses an inner mechanism of fast 
thermalization of the electronic excited states, for example,  
by emission of phonons \cite{thermo}.    
%
\begin{figure}[p]
\begin{center}
\leavevmode
\epsfxsize = 290pt
\epsfysize = 125pt
\epsfbox{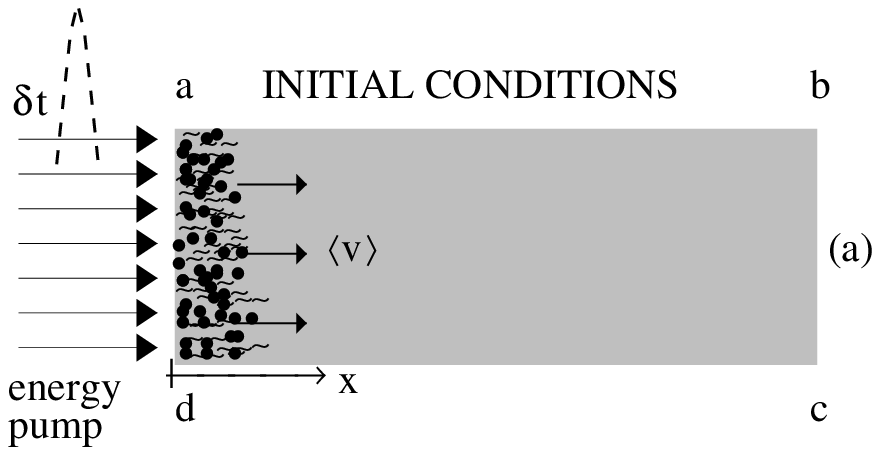}
\end{center}
\end{figure}
\begin{figure}[p]
\begin{center}
\leavevmode
\epsfxsize = 315pt
\epsfysize = 140pt
\epsfbox{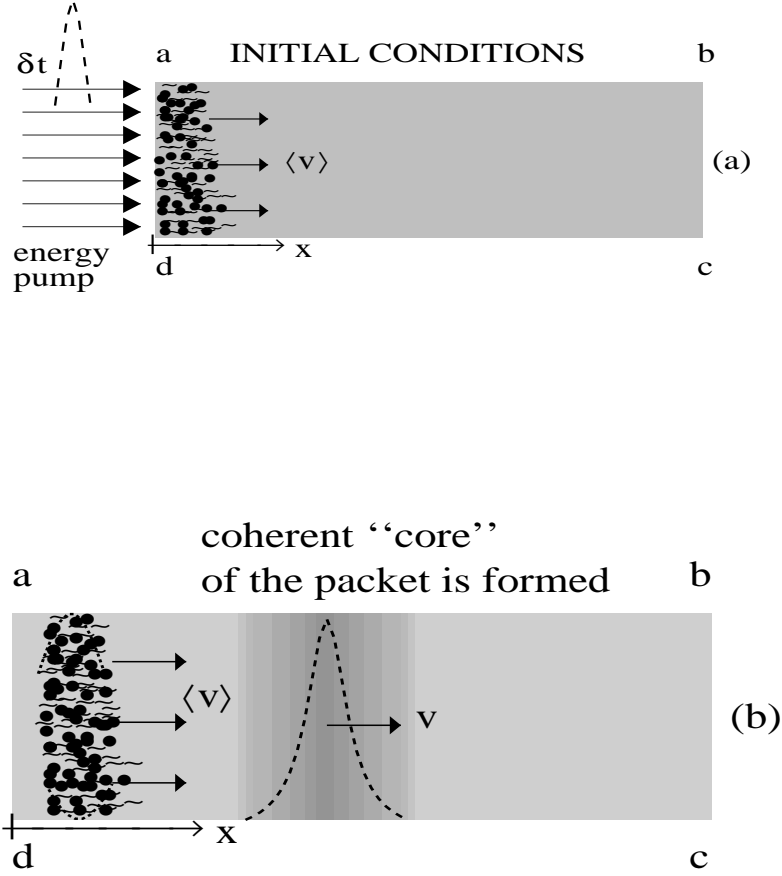}
\end{center}
\caption{ 
 A medium, in which the boson-phonon soliton can propagate, is presented on 
 Figures 1\,(a) and (b)
 in the form of  the channel `abcd'. 
 It has the dimensions $\vert {\rm ab}\vert=L$, \,
 $\vert {\rm bc}\vert \simeq \sqrt{S_{\perp}}$,
 and $L\gg L_{\rm ch}$, where  $L_{\rm ch}$ is the characteristic width of the soliton.
  After some amount of energy has been  pumped into the medium 
  during a short time interval $\delta t$ and absorbed near the boundary,
  a localized excited state is formed near the face `ad'.
  It is schematically shown on Figure 1\,(a) as a mixture of excitons and phonons.  
  If there is a 
  mechanism of the momentum transfer to the excited state, 
  the droplet begins to move toward
  the opposite face `bc' with the velocity $\langle v\rangle$, see Fig. 1\,(b). 
  Such conditions can favor the appearance of a coherent boson-phonon
  state (an analog of Davydov soliton) 
  moving ballistically along the axis $Ox$ at $T<T_{c}$.
  In other words, a sort of Bose-condensate can appear because of the effective attraction
  among the bosons (excitons) at $T<T_{c}$, see Fig. 1\,(b). The 
  coherent state, however, 
  is only a core of the total moving packet.
  The profile of the excitonic Bose-core,
  $n_{\rm o}(x,t) \simeq \vert \Psi_{\rm o}(x,t) \vert^{2}$, 
  is shown by the dashed line and the intensity of the elastic (phonon) part of the Bose-core,
  $\partial_{x}u_{{\rm o},\,x}(x,t)$, is represented  
  by changements of the intensity of the background color on Fig. 1\,(b).
}
\label{in_stage}
\end{figure}

It is known
\cite{elastic}
that  nonlinear elastic lattices support several types of  localized excitations;
some  of them turn out  to have  the so-called  nonzero  direct current, 
$
{ u}_{x} \simeq u_{\rm o}(x-\bar{v}t).
$
For example, a  moving packet  can consist of two parts, such as   
$$
\hat{ u}_{x} \sim u_{\rm o}(x-\bar{v}t) + 
\sum_{k} C(x-\bar{v}t)\,{\rm e}^{i(kx - \omega_{k}t )}\,\hat{b}_{k} +{\rm h.c.},
$$
and  $\bar{v}$ is the group velocity of the packet. 
On the other hand, the Bose-gas has a property of multiple occupancy, 
so that the coherent mode
can be introduced through the substitute \,
$\hat{\psi}(x,t) \rightarrow {\Psi}_{\rm o}(x,t)$  and 
\,$\int d{\bf x} \,\vert {\Psi}_{\rm o}(x,t)\vert ^{2} \gg 1$.
Thus, in the limit of $T\rightarrow 0$ and within the long 
wavelength approximation, 
one can explore how  Eqs. (\ref{eq1M}), (\ref{eq2M})
support the localized {\it  coherent} solutions. 
In this article, 
we use 
the language and technique of Bose-Einstein 
condensation  \cite{Griffin} that are suitable to take into account  
both the effect of many-particle coherency and the effect of 
many-particle nonlinearity.
Note that the two parts 
of the coherent field, 
${\Psi}_{\rm o} (x,t)$ and $u_{\rm o}(x,t)$, are taken into 
account selfconsistently and   
we do not integrate out the phonons.

Substituting the coherent fields ($c$-functions) into  operator equations
(\ref{eq1M}) and (\ref{eq2M}), 
we obtain the following system 
of dynamic equations 
\, ($\vartheta_{0}=( \hbar^{2}/2m)\,\bar{\vartheta}_{0}$ and  
$\bar{\vartheta}_{0}$ is dimensionless):
$$
i \hbar\partial_{t}\,{\Psi}_{\rm o}({ x},t)\,= 
\,-
(\hbar^{2}/2m)\,\bar{\vartheta}_{0}\,\partial_{x}{u}_{\rm o}(x,t)\,\partial_{x}^{2}{\Psi}_{\rm o}({ x},t)
\,+
$$
\begin{equation}
+\,\Bigr(\tilde{E}_{g} - {{\hbar^{2}}\over{2m}}\partial_{x}^{2} 
+ \nu_{0}
\vert \Psi_{\rm o}\vert ^{2}(x,t)+
\nu_{1}\vert  {\Psi}_{\rm o} \vert^{4}({ x},t)\,\Bigl)
{\Psi}_{\rm o}({ x},t) + \sigma_{0}\,\partial_{x} { u}_{\rm o} ( x,t)\,
{\Psi}_{\rm o}({ x},t),
\label{again1}
\end{equation}
$$
\partial_{t}^{2} { u}_{\rm o}({ x},t) - c_{l}^{2}\partial_{x}^{2} { u}_{\rm o}({ x},t) -
 c_{l}^{2}
2 \kappa_{3}\,\partial_{x}^{2} { u}_{\rm o}\,\partial_{x} { u}_{\rm o}({ x},t ) 
\,=
$$
\begin{equation}
=\,
\rho^{-1}\sigma_{0}\,\partial_{x}\bigl({\Psi}^{*}_{\rm o}{\Psi}_{\rm o}({ x},t)\bigr)
+ \rho^{-1} \vartheta_{0}\,\partial_{x}
\Bigl(\partial_{x}{\Psi}^{*}_{\rm o}\,\partial_{x}{\Psi}_{\rm o}({ x},t)\Bigr).
\label{again}
\end{equation}
This system can be considered as a generalization of 
the Zakharov system which appeared in Plasma Physics \cite{Zakharov}. 
The main difference is that, in the case of collective phenomena in 
Condensed Matter Physics, 
Eqs. (\ref{again1}) and (\ref{again}) define 
the main part of the moving packet. 
(It is called the coherent Bose-core in this article.)    
In fact, in the case of $T\ne 0$ and, generally, in all the nonstationary cases, 
these two equations have to be coupled with another system of equations on 
the so-called out-of-condensate excitations \cite{Griffin}. The dynamics 
of these excitation states can  strongly influence 
the dynamics of the ``parent'' condensate.

In this article, we rely on a kind of adiabatic hypothesis to 
simplify the solution of the problem.
If the occupancy of the coherent mode, 
$N_{\rm o}(t) = \int d{\bf x} \,\vert {\Psi}_{\rm o}(x,t)\vert ^{2}$, and the occupancy of the
out-of-condensate cloud  $\delta N (t) = 
\int d{\bf x} \,\langle \delta{\psi}^{\dag}\delta{\psi}(x,t)\rangle $,
change in time slowly and $N_{\rm o}(t)\gg \delta N(t)$, 
one can try to find some quasi-stationary solution for the coherent core 
${\Psi}_{\rm o}(x,t)$ 
and balance its energy and momentum by a kind of ``leakage'' 
from the Bose-core into the incoherent out-of-condensate cloud  and
tail, $N_{\rm o} \rightarrow N_{\rm o}-\delta N(t)$. 
This approach can be called the propagation of 
an exciton-phonon ``comet'' in a periodic medium, and it is valid within 
a finite time interval which is estimated below.

Recall that the standard ballistic ansatz  with $v={\rm const}$, 
$$
u_{\rm o}(x,t)=u_{\rm o}\left(x- t\, v\,\right),
$$
$$
{\Psi}_{\rm o}({ x},t)=\exp\bigl(i (\varphi_{\rm c} +  k_{0}x)\bigr) 
\,\exp(-i\Omega(k_{0})\,t)\,{\psi}_{\rm o}\!\left(
{ x}-t\, v\,\right),
$$
where
$$
\hbar \Omega(k_{0}) = 
\tilde{E}_{g} + 
(\hbar^{2}k_{0}^{2}/2m)
+\mu,\,\,\,\,\,\,\,\hbar \,k_{0}=m v,
$$
does not work properly in the case of \,$\vartheta_{0} \ne 0$.
Moreover, with such a ballistic ansatz,\, we always have 
\,$v= v_{\rm s}$,
where 
\,
$v_{{\rm s},\,x} \propto \partial_{x}\,\varphi_{\rm c}(x,t)={\rm
const}$ is the superfluid velocity of the Bose-core.
As the anomalous ballistic transport of exciton-phonon packets in
semiconductor and quasi-1D structures resembles, to some extent, 
the well-known effect of superfluidity \cite{Tilley},  
we apply the methods of Many-Particle Physics to Eqs. (\ref{again1}) and 
(\ref{again}).
We assume that the following substitutes for the ballistic velocity  
and the chemical potential, 
\begin{equation}
v \rightarrow v(x,t)\approx v\, 
\bigl(1 \pm \zeta(v)\,\psi_{\rm o}^{2}(x,t)+...\,\bigr), 
\label{NewBallisVelo}
\end{equation}
\begin{equation}
\mu \rightarrow \mu(x,t)\approx \mu\, 
\bigl(1 \pm \tilde{\zeta}(v)\,\psi_{\rm o}^{2}(x,t)+...\,\bigr), 
\label{NewChem}
\end{equation}
are more appropriate to model the nonlinear dynamics   
of the ballistic boson-phonon packets in the case of  
$\vartheta_{0} \ne 0$.  
However, within the model with Hamiltonian (\ref{twoT1}), 
one can  start, for example, from 
$$
v \rightarrow 
v(x,t)\approx v\,{\cal F}_{1}\bigl(\partial_{x}u_{\rm o}(x,t)\,\bigr)
$$
and {\it derive} expansions (\ref{NewBallisVelo}) and (\ref{NewChem}).

To simplify the dynamic equations, we choose the following ansatz: 
\begin{equation}
u_{\rm o}(x,t)=u_{\rm o}\Bigl(x- 
t\, v\,{\cal F}_{1}\bigl(\,\partial_{x}u_{\rm o}(x,t) \,\bigr)\Bigr),
\label{A_Z1}
\end{equation}
\begin{equation}
{\Psi}_{\rm o}({ x},t)=\exp\Bigl(\,i\varphi_{\rm c} + 
i\,k_{0}\,{\cal F}_{2}\bigl(\partial_{x}u_{\rm o}\bigr)\,
x\,-\,i\,\Omega(k_{0},\,\partial_{x}u_{\rm
o})\,t\Bigr)\,{\psi}_{\rm o}\Bigl(
{ x}-t\, v
\,{\cal F}_{1}\bigl(\,\partial_{x}u_{\rm o} (x,t)\,\bigr)\Bigr), 
\label{A_Z2}
\end{equation}
where  
$$
\hbar \Omega(k_{0},\,\partial_{x}u_{\rm o})\approx 
\tilde{E}_{g} + 
(\hbar^{2}k_{0}^{2}/2m)\,{\cal F}_{3}\bigl(\,\partial_{x}u_{\rm o} (x,t)\,\bigr)
+ \mu\,{\cal F}_{4}\bigl(\,\partial_{x}u_{\rm o} (x,t)\,\bigr).
$$
For ${\cal F}_{j}$, we use the following expansion: 
$$
{\cal F}_{j} \approx 1 + c_{j}\,\bar{\vartheta}_{0}\,\partial_{x}u_{\rm o}(x,t)
+ \bar{c}_{j}\,\bigl( \,\bar{\vartheta}_{0}\,\partial_{x}u_{\rm o}
(x,t)\,\bigr)^{2},
\,\,\,\, j=1,...,4,
$$
and $\hbar \,k_{0}=m v$. 
Note that, in some sense, ansatz (\ref{A_Z2}) is the well-known
representation 
$$
{\Psi}_{\rm o}({ x},t)=\sqrt{n_{\rm o}(x,t)}\,
\exp\bigl(i\varphi_{\rm c}(x,t)\,\bigr), 
$$
with the inhomogeneous fields \,$n_{\rm o}(x,t)$, 
$v_{{\rm s},\,x}(x,t) \propto \partial_{x}\varphi_{\rm c}(x,t)$, and \,
$\mu(x,t)\propto \partial_{t}\varphi_{\rm c}(x,t)$.
In this article, we will not write out and solve 
a set of nonlinear hydrodynamic equations \cite{hydro}
on these variables. 
Instead, we are interested in 
a microscopic approach aimed to clarify the role of
the coupling parameters, $\vartheta_{0}$
and $\sigma_{0}$.
Obviously, for a packet moving nonuniformly, 
one might expect some deviation from 
the simple low 
$$ 
\varphi_{\rm c}({\bf x},t)= \varphi_{\rm c} +  {\bf p}_{s}{\bf x}/\hbar -
(\tilde{E}_{g} + 
{\bf p}_{\rm s}^{2}/2m
+ \mu)\,t/\hbar.
$$
If a good initial approximation for such a  deviation 
can be found,  
one can finish with a kind of relatively simple NLS 
%
%
equation on the envelope functions of the Bose-core, 
$\exp(i\varphi_{\rm c})\,{\psi}_{\rm o}(x)$ and $\partial_{x}u_{\rm o}(x)$. 
Such a mathematical structure 
do appears from Eqs. (\ref{again1}) and (\ref{again}) within the validity of the adiabatic hypothesis
if one starts from  
quasistationary ansatz (\ref{A_Z1}) and (\ref{A_Z2}).
 
Recall that the  following relation between 
${\psi}_{\rm o}(x)$ and $\partial_{x}u_{\rm o}(x)$
is a  good approximation 
for the selfconsistent 
boson-phonon coherent state at $\vartheta_{0}=0$, 
$$
\partial_{x}u_{\rm o}(x,t) = f(\,{\psi}_{\rm o}^{2}(x,t)\,) 
\approx A_{2}\,{\psi}_{\rm o}^{2}(x,t) +
A_{4}\,{\psi}_{\rm o}^{4}(x,t).
$$ 
In fact, this approximation is valid in a more general case 
(although it is a part of a more complicated relation, such as\,  
$\partial_{x}u_{\rm o} = f({\psi}_{\rm o},\,\partial_{x}{\psi}_{\rm o})$\,).  
Then, for the coherent phase $\varphi_{\rm c}(x,t)$, 
we can write out the following symmetric representation:
$$
\varphi_{\rm c}(x,t)
\approx 
\varphi_{\rm c} +  k_{0}\,
\bigl\{1 + \zeta_{2}(v)\,\psi_{\rm o}^{2}(x,t) + \cdots \,\bigr\}\,x
\,- 
$$
\begin{equation}
-\,
\Bigl(\,\tilde{E}_{g} \,+\, 
\hbar k_{0}^{2}/2m\,\{1 + \zeta_{3}(v)\,\psi_{\rm o}^{2}(x,t) + \cdots\}
\,-\, \mu\,\{1 + \zeta_{4}(v)\,\psi_{\rm o}^{2}(x,t) + \cdots\}
\,\Bigr)\,t/\hbar. 
\label{phase_new}
\end{equation}
The ballistic velocity 
of the condensate, $x \rightarrow x-v(x,t)t$, 
is modified as follows:
\begin{equation}
x \,\rightarrow \,
x-v\bigl\{1 + {\zeta}_{1}(v)\,\psi_{\rm o}^{2}(x,t) + 
\cdots \,\bigr\}\,t.
\label{velo_new}
\end{equation}
Here, we introduce the parameters  $\zeta_{j}(v)\simeq c_{j}\zeta(v) \propto
\vartheta_{0}$ and expect
$\zeta_{1}(v)\simeq \zeta_{2}(v)\simeq \zeta_{3}(v)\simeq \zeta_{4}(v)
$.
If $c_{2}\simeq 1$, 
we get immediately  
$c_{1}=1+c_{2}\simeq 2$\, and \, $c_{3}=1+2c_{2}\simeq 3$
within the quasistationary approximation for the dynamic equations.
The next field terms (\,$\propto \bar{c}_{j}\,\vartheta_{0}^{2}\,\psi_{\rm
o}^{4}(x,t)$\,) 
in expansions 
(\ref{phase_new}), (\ref{velo_new}) 
can be defined through \,$c_{2}$ and $\bar{c}_{2}$ as well. 
However, they are small 
in a weakly nonlinear case and we will disregard them.
The constant $c_{4}$ will be defined to simplify the 
quasistationary equation on ${\psi}_{\rm o}^{2}(x)$.

To clarify the dynamic properties of the generalized ballistic { ansatz} and define the
unknown $c_{j}$, 
we will calculate 
the energy and 
momentum of a moving soliton-like solution with the factors \,${\cal
F}_{j}(x,t)$. 
Meanwhile, the envelope functions, 
$\exp(i\varphi_{\rm c})\,\psi_{\rm o}(x)$ \,and\, $\partial_{x}u_{\rm o}(x)$,
are found from the quasistationary equations one has to derive. 
These equations read 
$$
\mu\,\bigl\{1+ c_{4}\,\bar{\vartheta}_{0}\,\partial_{x}{u}_{\rm o}(x,t)+\cdots\,\bigr\}\, 
{\psi}_{\rm o}({ x},t) \approx
-
(\hbar^{2}/2m)\,\bar{\vartheta}_{0}\,\partial_{x}{u}_{\rm o}(x,t)\,\partial_{x}^{2}{\psi}_{\rm o}({ x},t)
\,+
$$
\begin{equation}
+\,\Bigr( - {{\hbar^{2}}\over{2m}}\partial_{x}^{2} 
+ \nu_{0}\,
 \psi_{\rm o}^{2}(x,t)+
\nu_{1}\,{\psi}_{\rm o}^{4}({ x},t)\,\Bigl)
{\psi}_{\rm o}({ x},t) + \sigma_{0}\,\partial_{x} { u}_{\rm o} ( x,t)\,
{\psi}_{\rm o}({ x},t),
\label{again11}
\end{equation}
$$
v^{2}\,\partial_{x}^{2} { u}_{\rm o}({ x},t)
\left(1 + c_{1}\,\bar{\vartheta}_{0}\,\partial_{x}u_{\rm o}
+\cdots
\,\right)^{2} - c_{l}^{2}\,\partial_{x}^{2} { u}_{\rm o}({ x},t)\,- 
$$
$$
-\,
 c_{l}^{2} \left(
2 \kappa_{3}\,\partial_{x}^{2} { u}_{\rm o}\,\partial_{x} { u}_{\rm o}({ x},t ) +
3 \kappa_{4}\,\partial_{x}^{2} { u}_{x}\,\bigl(\partial_{x} { u}_{\rm o}({ x},t)\bigr)^{2}\,
\right)\,\approx
$$
\begin{equation}
\approx \,
\rho^{-1}\sigma_{0}\,\partial_{x}\bigl({\psi}_{\rm o}^{2}({ x},t)\bigr)
+ \rho^{-1}\vartheta_{0}\,\partial_{x}
\Bigl({\partial_{x}\Psi}^{*}_{\rm o}\,\partial_{x}{\Psi}_{\rm o}({ x},t)\Bigr).
\label{again22}
\end{equation}
Note that we did not include the terms depending on time 
explicitly into these equations. 
The consequent divergencies, however, can be cured by 
taking into account the incoherent parts of the packet.
In this article, 
we propose the effect of slow dynamic redistribution of 
the occupation numbers between the condensate and 
out-of-condensate cloud, while both of these states can be easily calculated 
in the quasistationary approximation. 
Recall that   
dynamic mean field interaction between the condensate and 
noncondensed part of the packet leads to the damping of both of them
\cite{damping}.
On the other hand, the time interval, during which the exciton-phonon 
packet with the coherent Bose-core moves in the periodic medium, is always 
finite, especially for the ``clean'' samples without lattice
imperfections. Therefore, 
the adiabatic hypothesis can be used in our case. 

To find a soliton-like solution of Eqs. (\ref{again11}) and (\ref{again22}),
we reduce Eq. (\ref{again22}) to the following 
form:
$$
- (c_{l}^{2}-v^{2})\,\partial_{x}^{2} { u}_{\rm o}({ x},t) - 
 c_{l}^{2}
2\, \bar{\kappa}_{3}\,\partial_{x}^{2} { u}_{\rm o}\,
\partial_{x} { u}_{\rm o}({ x},t )
\,=
$$
\begin{equation}
=\,
\rho^{-1}\bar{\sigma}_{0}\,\partial_{x}\bigl({\psi}_{\rm o}^{2}({ x},t)\bigr).
\label{again33} 
\end{equation} 
Note that the interaction vertices $\kappa_{3}$ and $\sigma_{0}$  
are slightly renormalized in this equation,
$\bar{\kappa}_{3}=\bar{\kappa}_{3}(\vartheta_{0})$, \,$\bar{\sigma}_{0}=\bar{\sigma}_{0}(\vartheta_{0})$.
For example, 
we can write $\bar{\kappa}_{3}$ as follows 
($c_{2} \simeq 1$, $c_{1} \simeq 2$) 
\begin{equation}
\bar{\kappa}_{3} = -\vert \kappa_{3}\vert + 2\,(v^{2}/c_{l}^{2})\,
\vert \bar{\vartheta}_{0}\vert \,<\, 0,
\label{KappA3}
\end{equation}
and 
$\vert \bar{\kappa}_{3} \vert <   \vert \kappa_{3}\vert $ for 
$\bar{\vartheta}_{0}< 0$.
Here, we choose $\kappa_{3}<0$ ($\kappa_{4}>0$)
and assume the inequality \,
$$
\vert \kappa_{3}\vert  > \vert \bar{\vartheta}_{0}\vert  
$$
to be valid. Then, the sign and   
the order of value of \,$\kappa_{3}$\, 
remains the same after the renormalization. 
For example, we take
$\vert \kappa_{3}\vert \simeq 5 - 10$, whereas \,
$\vert \bar{\vartheta}_{0}\vert \simeq 1-2$,
and, for $v<c_{l}$, we can estimate \,$v^{2}/c_{l}^{2}\simeq 0.5 - 0.9$. 

One can represent the solution of Eq. (\ref{again33}) in the following form:
$$
\partial_{x}{ u}_{\rm o}({ x},t) \approx 
- A_{2}\,{\psi}_{\rm o}^{2}({ x},t) + 
A_{4}\,{\psi}_{\rm o}^{4}({ x},t), 
$$
where \,
$$
A_{2} \approx  \bar{\sigma}_{0}/\rho(c_{l}^{2} - v^{2})
=  \gamma(v)\,\Bigl(\bar{\sigma}_{0} /M c_{l}^{2} \Bigr)\,a_{l}^{3} > 0,
$$
$$
A_{4} \approx \gamma(v)\,\vert\bar{\kappa}_{3}\vert \,A_{2}^{2} > 0,
$$
and $\rho \simeq M /a_{l}^{3}$, \,$\gamma(v)=c_{l}^{2}/(c_{l}^{2} - v^{2}) > 1$.  
Immediately, we can rewrite the  factor ${\cal F}$
for the ballistic velocity 
($\bar{\vartheta}_{0}<0$) as follows:
$$ 
{\cal F}_{1}(\partial_{x}u_{\rm o}(x,t)\,) \rightarrow 
{\cal F}_{1}\bigl( \vert\Psi_{\rm o}(x,t) \vert^{2} \bigr )
 \,\approx
$$
\begin{equation}
\approx\,
1 + \vert \zeta_{1}(v)\vert \,{\psi}_{\rm o}^{2}({ x},t)
- \bar{\zeta}_{1}(v)\, {\psi}_{\rm o}^{4}({ x},t).
\label{U2}
\end{equation}
For $\bar{\vartheta}_{0}>0$, we estimate
$$ 
{\cal F}_{1}(\partial_{x}u_{\rm o}(x,t)\,) \rightarrow 
{\cal F}_{1}\bigl( \vert\Psi_{\rm o}(x,t) \vert^{2} \bigr )\, \approx 
$$
\begin{equation}
\approx \,
1 - \vert \zeta_{1}(v)\vert \,{\psi}_{\rm o}^{2}({ x},t)
+ \bar{\zeta}_{1}(v)\, {\psi}_{\rm o}^{4}({ x},t).
\label{U2U2}
\end{equation}

We use the following representation for $\vert\Psi_{\rm o}(x,t) \vert^{2}$:
$$
\vert \Psi_{\rm o}(x,t) \vert^{2} = {\psi}_{\rm o}^{2}(\,{ x}-v(x,t)t\,) \rightarrow 
\Phi_{\rm o}^{2}\,f^{2}(\bar{x}/L_{0}),  
$$
where $f(\bar{ x}/L_{0})$ is the dimensionless `shape' function, 
$\vert f({ x}/L_{0}) \vert \le 1$, \,$\Phi_{\rm o}$ is the amplitude of the coherent state, 
and 
${ x}-v(x,t)t  \rightarrow  \bar{x}$.
The characteristic width of the condensate is estimated as $(3-6)\,L_{0}$.
Although the terms \,$\propto \psi_{\rm o}^{4}$ in
Eqs. (\ref{U2}) and  (\ref{U2U2})
could be important in strong nonlinear regimes 
(as well as other omitted contributions, e.g.,  
\,$\propto (\partial_{x}\psi_{\rm o})^{2}$\,),
we assume 
\,$\zeta_{1}(v)\,{\Phi}_{\rm o}^{2}<1$ 
and 
\,$\zeta_{1}(v)\,{\Phi}_{\rm o}^{2} \gg \bar{\zeta}_{1}(v)\,{\Phi}_{\rm o}^{4}$
in this article.  
Then,  
we can roughly estimate the dynamic factors ${\cal F}_{j}$ as 
\begin{equation}
 {\cal F}_{j}\bigl( \,\vert\Psi_{\rm o}(x,t) \vert^{2} \,\bigr )
 \,\approx 1 \pm  c_{j}\,\zeta(v)\,{\Phi}_{\rm o}^{2}\,f^{2}({ x},t),
\,\,\,\,\,c_{j} \sim 1,
\label{FACT}
\end{equation}
and the question is how strong the factor ${\cal F}$ can deviate from 1.
If \,$\vert \bar{\vartheta}_{0} \vert \simeq  1$ \,
and \,$\gamma(v)\simeq 5 - 10$ ($v<c_{l}$),
\,we can estimate the dimensionless factor $c_{j}\zeta_{j}(v)/a_{l}^{3}$
that enters  
Eq. (\ref{FACT}) as follows  
$$
c_{j}\,\zeta_{j}(v)/a_{l}^{3} \simeq c_{j}\,\vert \bar{\vartheta}_{0} \vert \,\gamma(v)\,
\bigl (  \bar{\sigma}_{0}/Mc_{l}^{2} \bigr) \simeq 1. 
$$
Then, the most important parameter in Eq. (\ref{FACT})
remains to be \,$ a_{l}^{3}{\Phi}_{\rm o}^{2}\,<\,1 $.

Now, we can qualitatively describe the dynamics of bosonic 
core of the packet, see Fig. 2.
The top of the soliton \,$\vert \Psi_{\rm o}(\,x-v(x,t)t\,) \vert^{2} 
\rightarrow  \Phi_{\rm
o}^{2}\,f^{2}(\bar{x}/L_{0})$ moves with 
$$
v \,\rightarrow \,v \pm (\zeta_{1}(v)\,
{\Phi}_{\rm o}^{2})\,v, \,\,\,\,\,\, f_{\rm top}(x/L_{0}) \simeq 1,
$$ 
whereas the slopes 
of\,
$\vert \Psi_{\rm o}(x,t) \vert^{2} $ move  with 
$$
v \rightarrow v \pm 10^{-1}(\zeta_{1}(v)\,{\Phi}_{\rm o}^{2})\,v,\,\,\,\,\,\, 
f_{\rm slope}(x/L_{0}) \sim 10^{-1}. 
$$
The ballistic velocity of  
the tails of $\vert \Psi_{\rm o}(x,t) \vert^{2} $ remains to be 
almost unchanged, i.e. $\,=v$. 
One can introduce the important parameter \,$\delta v_{\rm top}$ \,as 
$$
\delta v_{\rm top} \,\approx \, \pm\, c_{1}\,\vert \zeta(v)\vert\,
{\Phi}_{\rm o}^{2}\,v
$$
and the dimensionless ratio
\begin{equation}
\frac{ \delta v_{\rm top}\,t}{L_{\rm o}} \,\approx \, \pm\,
c_{1}\,\vert \zeta(v) \vert\,
{\Phi}_{\rm o}^{2}\,vt\,\Big/\,L_{\rm o}. 
\label{ups}
\end{equation}
Then, the time scale $\Delta t$, during which our quasistationary ansatz can be
used to describe
the dynamics of the core of the total solitonic packets, can be roughly estimated from\, 
$\vert\delta v_{\rm top}\vert\,\Delta t\,/\,L_{\rm o}\,\sim\,1$.  
In fact, parameter (\ref{ups}) depends on time because the 
dynamic leakage 
from the Bose-core into the tail
and coma of the ``comet'' leads to ${\Phi}_{\rm o}(t)$ and $L_{\rm o}(t)$.
To estimate the values of $\vert \mu \vert$, $a_{l}^{3}\,{\Phi}_{\rm o}^{2}$, and $L_{0}$, 
we have to solve the stationary equation on \,${\psi}_{\rm o}({ x})$.

\section{Solution of stationary equation} 

As a result of all the simplifications we made, we obtain this equation in the
following form (we choose $c_{4}=1$): 
 $$
\mu\,\bigl\{
1+ {\zeta}(v)
\,\psi_{\rm o}^{2}(x) - \bar{\zeta}(v)
\,\psi_{\rm o}^{4}(x)
\bigr\}\, \psi_{\rm o}({ x})\,=
$$
\begin{equation}
=\,\Bigr(-
{{\hbar^{2}}\over{2m}}\Bigl\{\,1 +{\zeta}(v)
\,\psi_{\rm o}^{2}(x) - \bar{\zeta}(v)
\,\psi_{\rm o}^{4}(x)\,\Bigr\} 
\,\partial_{x}^{2}
- \vert \widetilde{\nu_0}(v)\vert \,
 \psi_{\rm o}^{2}(x)+
\widetilde{\nu_1}(v)\, \psi_{\rm o}^{4}({ x}) 
\,\Bigl)
{\psi}_{\rm o}({ x}).
\label{S_I_M_PLE}
\end{equation}
This equation resembles combined Ablowitz-Ladik and Nonlinear Schr\"odinger Equations
taken in the continuous limit \cite{Hennig},\cite{Salerno}.
Note that the boson interaction vertices are strongly renormalized because of
interaction with the coherent
phonon field $\partial_{x}u_{\rm o}$. 
We assume the following conditions to be valid  
\begin{equation}
\nu_{0}>0 \,\rightarrow \,\widetilde{\nu_0}(v/c_{l},\,\bar{\sigma}_{0})<0,\,\,\,\,\,\,\,
\nu_{1}>0\, \rightarrow \,\widetilde{\nu_1}(v/c_{l},\,\bar{\sigma}_{0},\,\bar{\kappa}_{3})>0.
\end{equation}

If the dimensionless parameter ${\zeta}(v)\,\Phi_{\rm o}^{2}
\propto a_{l}^{3}\,\Phi_{\rm o}^{2}$ 
is small enough, one can reduce Eq. (\ref{S_I_M_PLE}) 
to Nonlinear Schr\"odinger Equation. We use the representation
$$
 \psi_{\rm o}(x)= \Phi_{\rm o}\,
f \Bigl( \beta(\Phi_{\rm o}) x,\,\eta(\Phi_{\rm o})\Bigr),
\,\,\,\,\,
\mu = -\vert \mu\vert < 0,
$$
with unknown functions $\beta(\Phi_{\rm o}) \equiv L_{0}^{-1}$
and $\eta(\Phi_{\rm o})$. 
Then, the following simple equation appears as an approximation of  
Eq. (\ref{S_I_M_PLE}):  
\begin{equation}
 \vert \mu\vert \,f({ x})
- \vert\bar{\nu}_{0}(v)\vert \,\Phi_{\rm o}^{2}\,f^{3}({ x})
+ \bar{\nu}_{1}(v)\,\Phi_{\rm o}^{4}\,f^{5}({ x})
\approx
 {{\hbar^{2}}\over{2m}}\,\beta^{2}(\Phi_{\rm o})\,\partial_{x}^{2} f({ x}).
\label{N_S_E2_2}
\end{equation}
This equation is the so-called ``subcritical'' NLS equation. 
In Condensed Matter Physics, 
it is used in the theory of superfluidity and Bose-Einstein condensation 
\cite{Pomeau}. 
It is also applied in nonlinear optics for light pulses in the medium 
with a cubic-quintic
nonlinearity \cite{cub-quin}, and it is known 
as the Lienard equation in the theory of exactly solvable nonlinear equations 
\cite{Kong}. 

Here, we apply it to describe the coherent state of excitons and phonons 
with macroscopic occupancy. 
The following approximation is used for the renormalized vortices:  
\,$\bar{\nu}_{0}\approx \widetilde{\nu_{0}}$ \,and \,
$
 \bar{\nu}_{1}(v) \approx  \widetilde{\nu_1}(v)+  
\vert\widetilde{\nu_0}(v)\vert \,\zeta(v)$. 
To estimate their strength, 
we use the following formulas:
\begin{equation}
\vert \widetilde{\nu_ 0}(v) \vert = \gamma(v)\,\Bigl(\bar{\sigma}_{0} / M c_{l}^{2} \Bigr)\,
(\sigma_{0}\, a_{l}^{3} ) \,  - \,\nu_ {0} >0,
\end{equation}
which as valid at $\gamma(v) > \gamma_{\rm o}$, or, equivalently, 
$v_{\rm o}< v <c_{l}$ (for discussion, see \cite{Loutsenko},\cite{RL}), 
\begin{equation}
\widetilde{\nu_ 1}(v) = 
(\gamma(v)\,\vert \bar{ \kappa}_{3}\vert \,)
\Bigl\{ \gamma(v)\,\Bigl( \frac{ \bar{\sigma}_{0}}{ M c_{l}^{2}  } \Bigr)\, \Bigr\}^{2} 
\,(\sigma_{0}\, a_{l}^{6}) \,+\, \nu_ {1},
\end{equation} 
and 
\begin{equation}
 \bar{\nu}_{1}\,\Phi_{\rm o}^{4} \approx  \widetilde{\nu_1}\,\Phi_{\rm o}^{4} + 
 \vert\widetilde{\nu_0}\vert \,\Phi_{\rm o}^{2}\,
\Bigl({\zeta}(v) \,\Phi_{\rm o}^{2}\Bigr).
\end{equation}
Note that in  
the case  of $\vartheta_{0}<0$ we obtain the enhanced three 
particle vertex, and the nonlinear corrections are important to estimate its value,
$\nu_{1} \rightarrow \widetilde{\nu_1} \rightarrow 
\bar{\nu}_{1} > 0$.

The localized solution of Eq. (\ref{N_S_E2_2}) exists if 
the following inequalities  are valid \cite{Kong}:
\begin{equation}
\vert \mu\vert < \mu^{*}=\frac{3}{16}\,
\frac{ \vert\widetilde{\nu_ 0}\vert^{2}}{ \bar{\nu}_{1}},
\,\,\,\,\,\,\,
\Phi_{\rm o}^{2} < \Phi_{\rm o}^{*\,2}=\frac{3}{4}\,
\frac{ \vert\widetilde{\nu_ 0}\vert }{ \bar{\nu}_{1}}.
\end{equation}
Then, $\vert \mu\vert = \vert \mu\vert(\Phi_{\rm o})$ and we have 
$$
\vert \mu\vert/\mu^{*} =2\,\Phi_{\rm o}^{2}/\Phi_{\rm o}^{*\,2}\,-\,
\Bigl(\Phi_{\rm o}^{2}/\Phi_{\rm o}^{*\,2}\Bigr)^{2} < 1 .
$$
As a rough estimate, we can use the inverse formula  
$\Phi_{\rm o}^{2}\,(\mu) \approx 0.5\,(\,\vert \mu \vert / \mu^{*} )\,
\Phi_{\rm o}^{*\,2}$ \,valid \,at\, $\vert \mu \vert/\mu^{*} \ll 1$.
 
The representation \,$\vert \mu\vert = (\hbar^{2} / 2m)\,L_{0}^{-2}$ 
leads to an easy estimate of the characteristic width of the soliton. Indeed,  
we can introduce the length  $L_{*}$ through 
$\mu^{*} = (\hbar^{2} / 2m)\,L_{*}^{-2}$ 
\,and\, obtain the following
representation  
\begin{equation}
\beta (\Phi_{\rm o}) \,\rightarrow \,
\beta (\mu) = \sqrt{ \frac{2m}{\hbar^{2}}\,\mu^{*}\,
(\vert \mu\vert /\mu^{*}) },\,\,\,\,\, 
\beta (\,\vert\mu\vert\,)= L_{*}^{-1}\,\sqrt{\vert \mu\vert /\mu^{*} }.
\end{equation}
Then, \, $L_{0} = 
L_{*}\,\sqrt{ \mu^{*} / \vert \mu\vert } $ \,and, always, \,
$L_{0} > L_{*}$.
To estimate the important microscopic parameters  $\mu^{*}$ and  $L_{*}$,
we write the renormalized vertices in the following form:   
\begin{equation}
\vert\widetilde{\nu_ 0}(v)\vert =
\tilde{\alpha}_{0}(v)\,{\rm Ry}_{\rm x}\,{a}_{\rm x}^{3} 
\,\,\,\,{\rm and}\,\,\,\,\bar{\nu}_{1}(v)=
\bar{\alpha}_{1}(v)\,{\rm Ry}_{\rm x}\,{a}_{\rm x}^{6},
\end{equation}
and estimate\,
\,$\tilde{\alpha}_{0} \simeq \bar{\alpha}_{1}  \simeq 10^{-1}$. 
(In theory, $\tilde{\alpha}_{0}$ can vary from $\sim 10^{-2}$ to $\sim 1$ with  
 changement of  $v$ within \,$v_{\rm o} < v <c_{l}$ \cite{Loutsenko},\cite{RL}.)
Then, we have
\begin{equation}
\mu^{*}(v)= \frac{3}{16}\,
\frac{\tilde{\alpha}_{0}^{2}}{\tilde{\alpha}_{1}}\,{\rm Ry}_{\rm x}\simeq
10^{-2}\,{\rm Ry}_{\rm x}, 
\,\,\,\,\,\,\,
L_{*}^{2}(v) 
\simeq \frac{\tilde{\alpha}_{1}}{\tilde{\alpha}_{0}^{2}}\,{a}_{\rm
x}^{2} \sim 10\,{a}_{\rm x}^{2}.
\label{*parameters}
\end{equation}

As the following inequality is valid 
$$
\zeta_{j}(v) \,\Phi_{\rm o}^{2} \,< \, \zeta_{j}(v) \,\Phi_{\rm o}^{*\,2},
$$
we can define the upper limit of  \,
$\delta v_{\rm top}/v \simeq \zeta_{1}(v)\,\Phi_{\rm o}^{2}$
in our model. We have
\begin{equation}
\zeta_{1}(v) \,\Phi_{\rm o}^{*\,2}(v)
\approx 
\vert \bar{\vartheta} _{0}\vert \,c_{1}\,
\left\{ \gamma(v)\, \frac{ \bar{\sigma}_{0}}{ M c_{l}^{2}  } \right \} \,
\bigl(a_{l}^{3}\,\Phi_{\rm o}^{*\,2} \bigr)\,\simeq \,0.1 - 1, 
\label{MOVE}
\end{equation}
where
\begin{equation}
a_{l}^{3}\,\Phi_{\rm o}^{*\,2}(v) \simeq 
\bigr(\,\tilde{\alpha}_{0}(v)/\bar{\alpha}_{1}(v)\,\bigl)\,
(a_{l}^{3}/{a}_{\rm x}^{3})
\simeq 0.1. 
\end{equation}
As a result, one can define the meaning of the weakly nonlinear case
in our model. For example, 
if the parameter \,$\Phi_{\rm o}^{2}/\Phi_{\rm o}^{*\,2} < 0.1 - 0.5$, 
the assumption 
$$ 
\bar{\zeta}(v) \,\Phi_{\rm o}^{4}
\ll 
\zeta (v) \,\Phi_{\rm o}^{2}
$$
we made to simplify the factors 
${\cal F}_{j}(\,\partial_{x}u_{\rm o}(x,t)\,)$ 
\,  to\, ${\cal F}_{j}(\,\psi_{\rm o}^{2}({ x},t)\,)$
is correct. 

We write the exact solution of Eq. (\ref{N_S_E2_2}) in the following form: 
\begin{equation}
\psi_{\rm o}^{2}(x)=
 \frac{ \Phi_{\rm o}^{*\,2}\,\bigl(\vert\mu \vert /2\mu^{*}\bigr)}
{
\sqrt{1-\vert \mu \vert /\mu^{*} }\,
\cosh^{2}\bigl(\beta(\vert \mu \vert )\,x\bigr)\,+\,(1/2)(1-\sqrt{1-\vert \mu \vert /\mu^{*}})
}    
\label{SolitonMOD2}
\end{equation}
Note that
$$
\psi_{\rm o}^{2}(x=0)\,=\,\Phi_{\rm o}^{2}(\,\vert \mu \vert\,) = 
\Phi_{\rm o}^{*\,2}
\,(1-\sqrt{1-\vert \mu \vert /\mu^{*}}).
$$
If \,$\vert \mu  \vert \ll \mu^{*}$, we can use the following asymptotics
(\,$\eta(\vert \mu \vert /\mu^{*}) \rightarrow 0$):
\begin{equation}
\psi_{\rm o}(x) \simeq 2 \Phi_{\rm o}\, {\rm exp}(-\beta(\Phi_{\rm o})\,\vert x\vert\, )
\,\,\,\,\,\,{\rm at}\,\,\,\,\,\,\vert x \vert >  2\,\beta(\Phi_{\rm o})^{-1}.
\end{equation}
In addition, the coherent phonon part ${ u}_{\rm o}({ x},t)$ can be rewritten 
as follows
$$
\partial_{x}{ u}_{\rm o}(x) \approx 
-\gamma(v)\,\Bigl(\bar{\sigma}_{0} /M c_{l}^{2} \Bigr)\,
(a_{l}^{3}\,\Phi_{\rm o}^{*\,2} ) 
\,\Bigl(\Phi_{\rm o}^{2} / \Phi_{\rm o}^{*\,2}\Bigr)\,
f^{2}\bigl(\beta(\vert \mu \vert )\, x\bigr)
\,+ 
$$
\begin{equation}
+ \,\gamma(v)\, \vert \bar{\kappa}_{3}\vert
\,\Bigl\{
\gamma(v)\,\Bigl(\bar{\sigma}_{0} /M c_{l}^{2} \Bigr)\,
(a_{l}^{3}\,\Phi_{\rm o}^{*\,2} ) 
\Bigr\}^{2}
\Bigl(\Phi_{\rm o}^{2} / \Phi_{\rm o}^{*\,2}\Bigr)^{2}\,
f^{4}\bigl(\beta(\vert \mu \vert )\, x\bigr).
\label{DERIV}
\end{equation}

We apply the 3D normalization condition on the macroscopic 
wave function $\Psi_{0}(x,t)$ as follows
$$
\int\!\!\vert\Psi_{0}\vert^{2}(x,t)\,d{\bf x} =N_{\rm o}(t) \gg 1.
$$
Substituting the quasistationary ansatz into the integrand, we have\, 
\begin{equation}
S\int\!\!\psi_{\rm o}^{2}\Bigl(x-v(1+\zeta_{1}(v)\psi_{\rm o}^{2}(x,t)+...)t\,
\Bigr)\,dx = N_{\rm o}(t)\,\rightarrow 
\end{equation} 
\begin{equation}
\rightarrow \,S\int\!\!\psi_{\rm o}^{2}(x')\,dx' = N_{\rm o}.
\label{norma}   
\end{equation} 
Here, $N_{\rm o} \gg 1$ has a meaning of the number of particles inside the
quasistationary condensate,
(it is a macroscopic number), 
and $S\approx L_{\perp}^{2} \simeq S_{\perp}$, where 
$S_{\perp}$ is the cross section area of the crystal. Note that, due to the
symmetry of $f(x/L_{0})$ against $x\rightarrow -x$, the value of 
$N_{\rm o}$ is conserved, but as a first approximation.

To understand  how  the macroscopic wave function is formed 
from the microscopic parameters of the theory
(they are $\mu^{*}$, $\Phi_{\rm o}^{*}$, $L_{*}$, and
the renormalized vertices $\tilde{\nu}(v)$), 
it is useful to introduce the parameter \,
$N^{*}=S\,L_{*}\,\Phi_{\rm o}^{*\,2}$ \,as well.
We estimate  $N^{*}$ as  follows:
\begin{equation}
N^{*}(v) \simeq \Bigl(1/\sqrt{2\, \bar{\alpha}_{1}(v)} \Bigr)\,
(S / {a}_{\rm x}^{2}) \gg 1.
\end{equation}
Then,\, $N^{*}(v) \sim  (S / {a}_{\rm x}^{2}) $ is 
always the  macroscopic number in 3D case
($\bar{\alpha}_{1}(v) \simeq 10^{-1} - 10^{-2}$). 
In fact,  
Eq. (\ref{norma}) leads to an algebraic equation that relates the macroscopic parameters with 
the microscopic ones. As a result,
\,  
$\sqrt{\vert \mu \vert/\mu^{*}}$ 
can be defined as a function of $\exp(2\,N_{\rm o}/N^{*})$. To avoid cumbersome
formulas, we use simple estimates: 
\begin{equation}
\sqrt{\vert\mu\vert/\mu^{*}}= f(\,N_{\rm o}/N^{*}\,) \simeq N_{\rm o}/N^{*}\,\,\,\,\,{\rm and}\,\,\,\,\,
L_{0}/L_{*}\simeq (N_{\rm o}/N^{*})^{-1}.
\end{equation} 
They are valid 
up to $N_{\rm o}/N^{*} \simeq 0.1 - 0.5$.

Thus, we can estimate the important dimensionless parameter that characterizes the 
deviation of  the ballistic velocity
$v(x,t)$ from $v={\rm const}$ and 
the chemical potential $\mu(x,t)$ from $\vert\mu\vert ={\rm const}$.
We can write
\begin{equation}
\Phi_{\rm o}^{2}/ \Phi_{\rm o}^{*\,2}= {\rm f}(\,\vert\mu\vert/\mu^{*}\,) \approx (1/2)(\vert \mu \vert
/\mu^{*})\simeq (1/2)(N_{\rm o}/N^{*})^{2}.
\end{equation}
This means 
$$
\Phi_{\rm o}^{2}/ \Phi_{\rm o}^{*\,2}\simeq   10^{-3} \rightarrow 10^{-2}
 \,\,\,\,\,{\rm while}\,\,\,\,\, N_{\rm o}/N^{*}(v) \simeq
10^{-2}\,\rightarrow\,10^{-1},
$$
and 
$$
\Phi_{\rm o}^{2}/ \Phi_{\rm o}^{*\,2}\simeq  0.2\,\rightarrow \,0.4
\,\,\,\,\,{\rm while}\,\,\,\,\, N_{\rm o}/N^{*}(v) \simeq 0.6-0.7 \,
\rightarrow \,1.
$$
As a result, 
the reasonable values for the macroscopic parameter $\zeta(v)\,\Phi_{\rm o}^{2} \simeq \delta
v_{\rm top}/v $ \,
can be estimated as
\begin{equation}
\bigl\{\,\zeta(v)\,\Phi_{\rm o}^{*\,2}\,\bigr\}
\,\Bigl ( \Phi_{\rm o}^{2} / \Phi_{\rm o}^{*\,2} \Bigr  ) 
\simeq 10^{-3}-10^{-2},
\label{howMANY}
\end{equation}
(the microscopic\, $\zeta(v)\Phi_{\rm o}^{*\,2} \sim 0.1-1$).
Now, 
we can rewrite parameter (\ref{ups}) to emphasize 
the interplay between the microscopic and macroscopic parameters of 
the theory: 
\begin{equation}
\frac{ \delta v_{\rm top}\,t}{L_{\rm o}} \,\simeq  \, 
\pm\, 0.5\,c_{1}\,\Bigl\{\,\vert \zeta(v)\vert\,{\Phi}_{\rm
o}^{*\,2}\,\Bigr\}\,
\frac{ vt}{L_{*}}\,
(N_{\rm o}/N_{*})^{3},\,
\,\,\,\,\,\, \zeta(v)\propto \vartheta_{0}\,\sigma_{0}.
\label{ups1}
\end{equation}
On Fig. 2, we show how the coherent part of the packet 
moves in the medium if one switch on the interaction \,$\vartheta_{0}\ne 0$.  
The initial state is taken to be symmetric 
(the exact solution for $\sigma_{0}\ne 0$, \,$\vartheta_{0}=0$ is used), and its evolution is 
obtained by use of quasistationary ansatz (\ref{A_Z1}), (\ref{A_Z2}) which 
conserves the amplitude and the characteristic width of the moving soliton.
Although it is a rough estimate of the real dynamics of the Bose-core,  
it gives some understanding of how the microscopic exciton-phonon interaction controls
the dynamics.  In addition, it can be easily adjusted to a 
more realistic case of the moving core and out-of-condensate cloud.

\newpage
\begin{figure}[p]
\begin{center}
\leavevmode
\epsfxsize = 350pt
\epsfysize = 200pt
\epsfbox{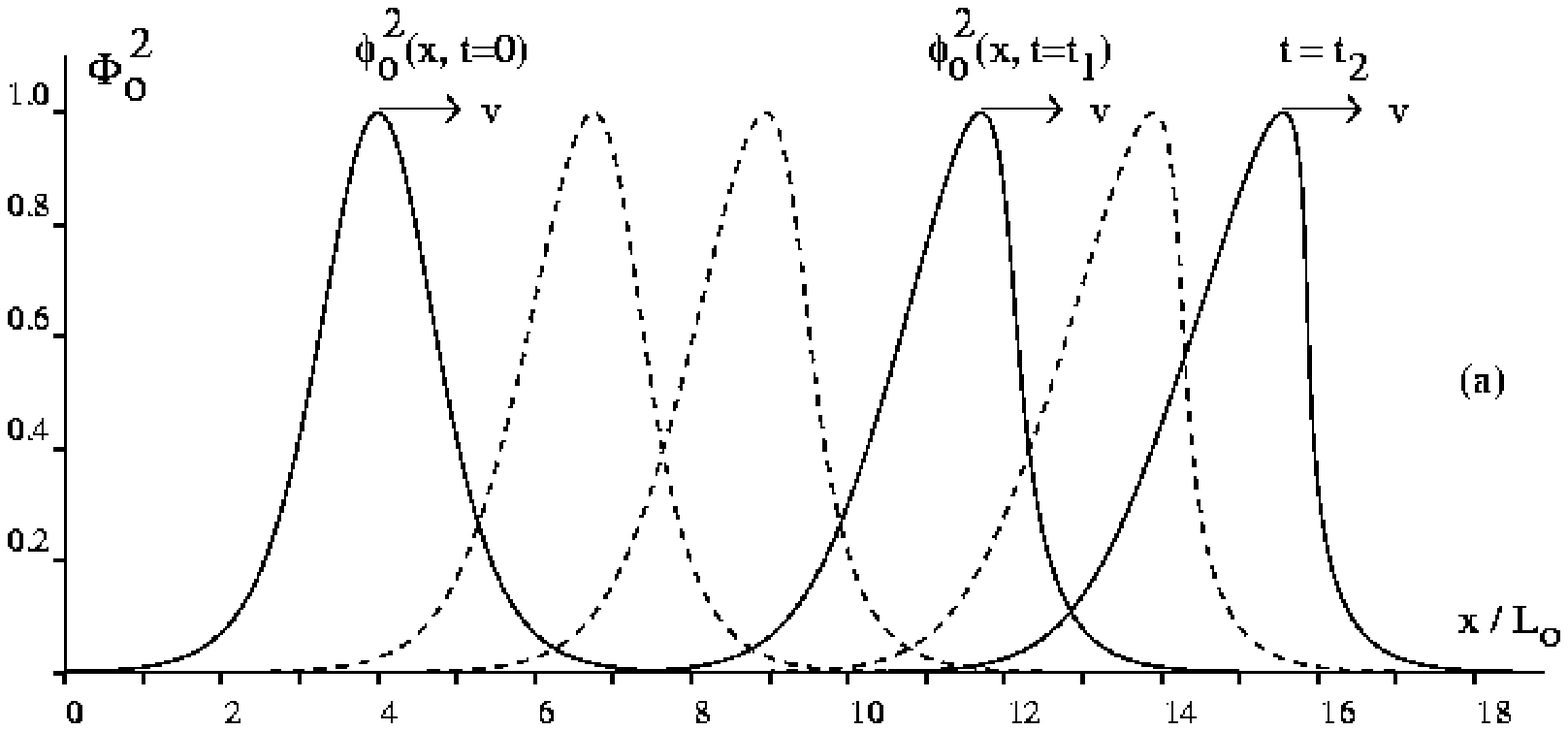}
\end{center}
\end{figure}
\begin{figure}[p]
\begin{center}
\leavevmode
\epsfxsize = 350pt
\epsfysize = 125pt
\epsfbox{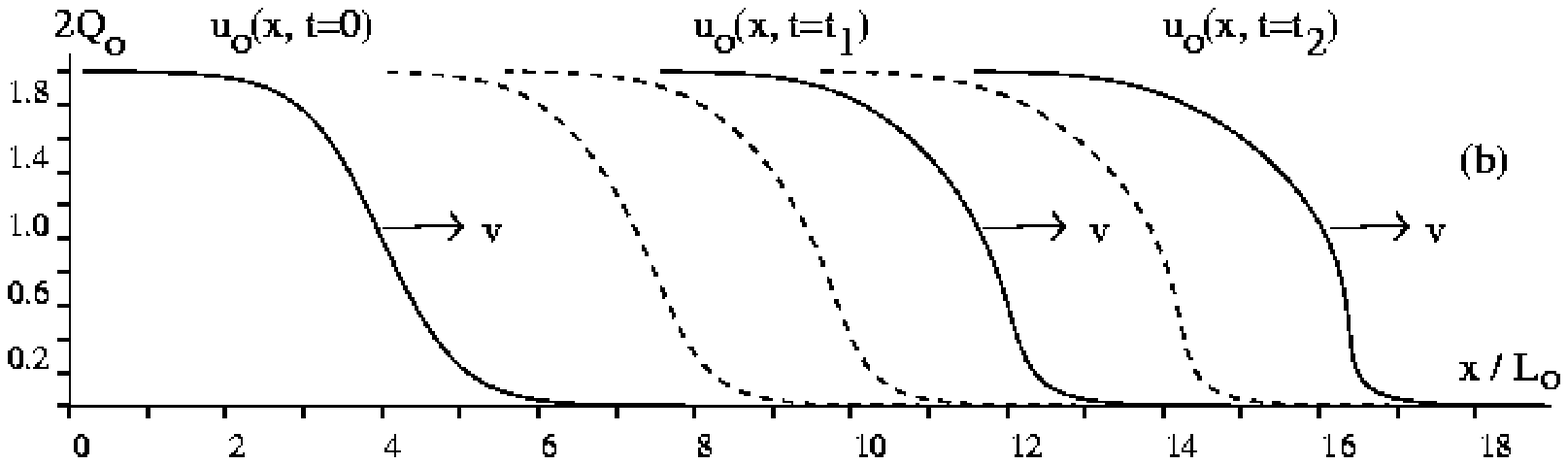}
\end{center}
\end{figure}
\begin{figure}[p]
\begin{center}
\leavevmode
\epsfxsize = 350pt
\epsfysize = 200pt
\epsfbox{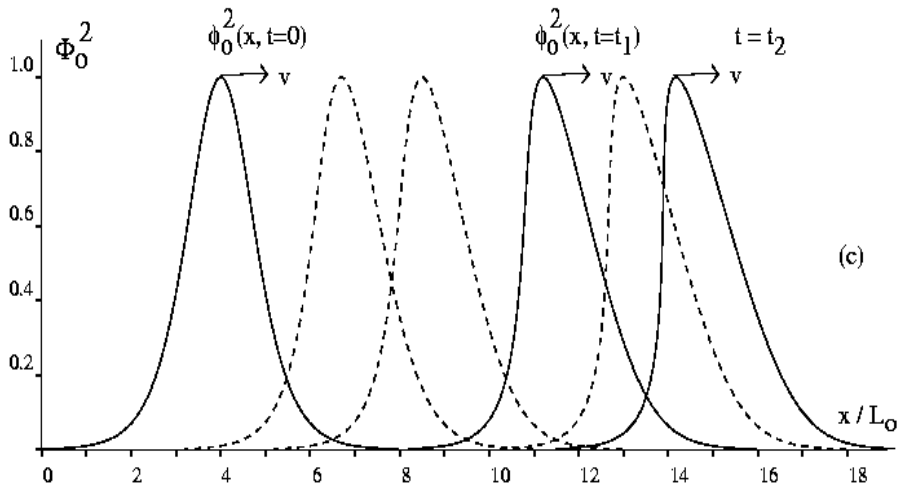}
\end{center}
\caption{
To model transport properties of the boson-phonon soliton in a periodic medium, 
we started from the symmetric soliton as an initial condition at $t=0$.
Dynamics of the boson (exciton) part of the packet is presented on 
Figs. 2\,(a) and 2\,(c) in form of the moving 
\,$\vert \Psi_{\rm o}(\,x-v(x,t)t\,)\vert ^{2}$ without any ``leakage''. 
The phonon part (a moving kink of the displacement field of the medium) 
is depicted on 
Fig. 2\,(b). (As a rough estimate,  
\,$\partial_{x}u_{\rm o}\bigl(x-v(x,t)t\bigr) \simeq - A_{2}\,\psi_{\rm
o}^{2}\bigl(x-v(x,t)t\bigr)$.\,)  
The interaction 
parameter $\zeta_{1}(v)\,\Phi_{\rm o}^{2} \simeq \delta v_{\rm top}/v$ 
controls the changement in the packet shape. It is taken to be 
$+0.1$ (Figs. 2\,(a) and (b)) and $-0.1$ 
(only the Boson part of the packet is presented on Fig. 2\,(c)). 
Then, the visible changements 
occur after the packet has traveled the distance of $n\, L_{0}$ corresponding to    
\,$ \delta v_{\rm top}\,\Delta t / L_{0} \simeq 0.5-1$.  
\,For \,$\delta v_{\rm top}/v=\pm 0.1$, we estimate \,$n \sim 10$.
Note that the presented result is a crude estimate 
of the dynamics of the Bose-core 
of the total moving packet.    
To proceed, 
one has to ``dress'' such a core with the out-of-condensate cloud and tail.  
}
\label{bare_soliton}
\end{figure}

\newpage

\section{Energy of the moving soliton}

Recall that the standard ballistic boson-phonon soliton, 
$\Psi_{\rm o}(x,t)= \exp\bigl(\,i \varphi_{\rm c}(x,t)\,  \bigr)
\,\phi_{\rm o}(x-vt)$ and $\partial_{x}u_{\rm o}(x-vt)$
with \, 
$\varphi_{\rm c}(x,t)=\varphi_{\rm o}+ k_{0}x -
\omega_{0}\,t$,
is a stationary state ($\vartheta_{0}=0$), \,that is   
$$
E_{\rm  o}=\int d{\bf x}\,{\cal T}_{0}^{0}(x,t)= {\rm const}
\,\,\,\,{\rm and}\,\,\,\,
{\bf P}_{\rm  o} \rightarrow  P_{{\rm  o},\,x}=
\int d{\bf x}\,{\cal P}_{x}(x,t) ={\rm const}. 
$$
At $\vartheta_{0}\ne 0$, we obtained an analog of such a solution,  starting from
Eqs. (\ref{A_Z1}), (\ref{A_Z2})  with   $v \rightarrow v(x,t)$, 
$k_{0}\rightarrow k_{0}(x,t)$, and $\mu\rightarrow \mu(x,t)$. 
In this article, the quasistationary 
approximation is the most important one we used  
to simplify the dynamic equations for the core of the packet.
Therefore, 
the exact calculation of $E_{\rm o}$ leads to 
$\partial_{t}E_{\rm  o}\ne  0$ \,for 
\,$\Psi_{\rm o}(x,t) =\exp\bigl(\,i\varphi_{\rm c}(\,x,t,\psi_{\rm o}(x,t)\, )\,\bigr) 
\,\psi_{\rm o}(\,x-v(x,t)t)$ with $N_{\rm o}={\rm const}$ and $L_{0}={\rm const}$.
Within the validity of the adiabatic hypothesis 
(slow exchange between $N_{\rm o}(t)$ and $\delta N (t)$ 
 while $N_{\rm o}(t) \gg \delta N (t)$\,),
one has, first, to calculate $\partial_{t}E_{\rm  o}$ of the quasistationary Bose-core
and, second,  to cure the obtained divergency by  
$\partial_{t}E_{\rm tot} = \partial_{t}E_{\rm  o} + \partial_{t}\delta E = 0$\,
($T\rightarrow 0$).


To calculate the energy of the moving condensate,  
we have to integrate the zero component
of the energy-momentum tensor ${\cal{T}}_{0}^{0}$ over the spatial coordinates.
We have the 
following formula (written in the laboratory frame): 
$$
{\cal{T}}_{0}^{0}({\bf x},t)= \tilde{E}_{g}\Psi^{*}_{\rm o} \Psi_{\rm o} + 
\frac{\hbar^{2}}{2m}\nabla\Psi^{*}_{\rm o}\,\nabla\Psi_{\rm o} 
+ \frac{\nu_{0}}{2}\left( \Psi^{*}_{\rm o} \right)^{2}\Psi_{\rm o}^{2}
+\frac{\nu_{1}}{3}\left( \Psi^{*}_{\rm o} \right)^{3}\Psi_{\rm o}^{3}
+
$$
$$
+ \,\frac{\rho}{2}(\partial_{t} u_{\rm o})^{2} +\frac{\rho c_{l}^{2}}{2}\,(\partial_{x} u_{\rm o})^{2}
+ \frac{\rho c_{l}^{2}}{3}\, \kappa_{3}\, (\partial_{x} u_{\rm o})^{3}
\,+\,
\sigma_{0}\,\partial_{x} u_{\rm o}\,
\Psi^{*}_{\rm o}\Psi_{\rm o}
\,+\,
\bar{\vartheta}_{0}\,\frac{\hbar^{2}}{2m}
\,\partial_{x} u_{\rm o}\,
\nabla\Psi^{*}_{\rm o}\,\nabla\Psi_{\rm o}.
$$
After integration of ${\cal{T}}_{0}^{0}$, we conclude that 
there are no terms $\propto t^{1}$ in $E_{\rm o}(t)$, 
\begin{equation}
E = E_{\rm o} + (\delta E)\,(vt/L_{0})^{2} + \cdots \,\approx\, 
e_{0}\,N_{\rm o}\, + \,\delta e_{0}\,N_{\rm o}\,
\bigl(\delta v_{\rm top}\,t\,/\,L_{0}\bigr)^{2}.
\label{with_t}
\end{equation} 
In fact, the dependence on $N_{\rm o}$ is a  nonlinear one, 
$$
E \approx 
e_{0}(N_{\rm o}/N_{*})\,N_{\rm o}\, + \,
\delta e_{0}(N_{\rm o}/N_{*})\,N_{\rm o}\,
\bigl\{\,\delta v_{\rm top}(N_{\rm o})\,t\,/\,L_{0}(N_{\rm
o})\,\bigr\}^{2}.
$$ 
We found \,$\delta e_{0}>0$ \,for\,  both $\vartheta_{0}>0$ 
\,($\delta v_{\rm top}<0$)\,
 and \, $\vartheta_{0}<0$ \,($\delta v_{\rm top}>0$). 
Indeed, we can write out the expansion
$$
\delta e_{0} \approx \delta e_{0}^{(0)}\, + \,
\delta e_{0}^{(1)}\,(\zeta(v)\,\Phi_{\rm o}^{2})\,+\,\cdots
\,\simeq \,\delta e_{0}^{(0)} \,>\,0, 
$$
and\,$\delta e_{0}^{(1)} \sim \delta e_{0}^{(0)}$.
Recall the structure of the stationary part, \,
$E_{\rm o}= e_{0}\,N_{\rm o}= E_{\rm x}+ E_{\rm int}+ E_{\rm ph}$.  We
have
\begin{equation}
e_{0}\approx (\hbar^{2} k_{0}^{2}/2m)\bigl(\,1+{\rm
const}\,\zeta(v)\,\Phi_{\rm o}^{2} + ...)
\, -\,{\rm const}'\,\vert \mu\vert \,+\,e_{\rm ph},
\label{e_e}
\end{equation}
where \,${\rm const},\,{\rm const}' \sim 1$ \,and \,
\,$e_{\rm ph}\approx  e_{\rm ph}^{(2)} - e_{\rm ph}^{(3)}$.
Here, the harmonic contribution into $E_{\rm ph}$ is \,
$e_{\rm ph}^{(2)}\,N_{\rm o} >0$, and
$$
e_{\rm ph}^{(2)}
\simeq \,\frac{M\,(v^{2}+c_{l}^{2})}{2}
\left(\gamma(v)\,\frac{\sigma_{0}}{ M c_{l}^{2} }\right)^{2}
(a_{l}^{3}\,\Phi^{*\,2}_{\rm o})\,\Bigl\{0.5\,\vert \mu\vert/\mu^{*}\Bigr\}.
$$ 
We can compare formula (\ref{e_e}) with our estimate of $\delta e_{0}^{(0)}$:
\begin{equation}
\delta e_{0} \simeq (\hbar^{2} \tilde{k}_{0}^{2}/2m)\,(c_{3}^{2}/4c_{1}^{2})
\,{\rm const}_{1}\,
+ \,\vert \mu\vert\,3\,{\rm const}_{2} + e_{\rm ph}^{(2)}\,{\rm const}_{3}
\,>\,0,
\label{dd-ee}
\end{equation}
where all the constants are small 
(as a rough estimate, \,${\rm const}_{j} \sim 10^{-1}$).
Then, \,$\delta e_{0} \ll e_{0}$.

The result (\ref{with_t}) means that 
changement of the shape of the Bose-core, which was included  
into the generalized ballistic  ansatz to satisfy 
the dynamic equations with $\vartheta_{0} \ne 0$,   
costs energy for both $\delta v_{\rm top}>0$ and 
$\delta v_{\rm top}<0$ during the observation time.
Note that 
the width of the soliton, $(3-6)\,L_{0}(N_{\rm o})$,
\,$L_{0}^{-2} \propto \vert \mu \vert(N_{\rm o})$,
and the number of Bose-particles forming the soliton, $N_{\rm o}$, 
are conserved as it was presented on  Fig. 2, but this is only  
the first approximation within the adiabatic hypothesis. 
It is possible, however, to balance the energy of the {\it total} boson-phonon
packet. 
(Here, the image of an exciton-phonon ``comet'' helps: 
the Bose-core with $N_{\rm o}\gg 1$ is only a part of the moving  
delocalized
packet.) 

First, we can calculate the value of \,
$\{ \,\partial_{N_{\rm o}}\,E_{\rm o}\,\}\,\delta N $.  
We take it as 
$$
\bigl\{
\partial_{N_{0}}\,E_{\rm o}
\bigr\}\,\delta N  \,\simeq \,\widetilde{ e_{0}}\,\delta N ,
$$
where we have the following formula
for $\widetilde{ e_{0}}$, ($\widetilde{ e_{0}} \ne  e_{0}$ in Eq.(\ref{e_e})\, ): 
\begin{equation}
\widetilde{e_{0}}\simeq (\hbar^{2} k_{0}^{2}/2m)\bigl(\,1+{\rm
const}_{\rm new}\,\zeta(v)\,\Phi_{\rm o}^{2} + ...)
\,-\,{\rm const}'_{\rm new}\,\vert \mu\vert \,+\,\tilde{e}_{\rm ph},
\label{dd-ee1}
\end{equation}
and\,
$\tilde{e}_{\rm ph}\simeq 3\,{e}_{\rm ph}^{(2)} - 5 \,{e}_{\rm ph}^{(3)}$.
\,Second, we assume that there is a  ``leakage'' from 
the moving localized state, and such a
leakage occurs dynamically,     
\begin{equation}
\delta N(t) \simeq - \delta N_{\rm o}(t) 
\simeq - \,{\rm const}\, N_{\rm o}\,\bigl(\delta v_{\rm top}\,t\,/\,L_{0}\bigr)^{2}
\,<\,0. 
\label{Leak}
\end{equation}
Then,
the tail behind the condensate  {\it grows} as 
\begin{equation}
\langle \delta \hat{N}_{\rm x, \,tail} \rangle \simeq  
 \delta N_{\rm o}(t) = {\rm const}'\, 
N_{\rm o}\,\bigl(\delta v_{\rm top}\,t\,/\,L_{0}\bigr)^{2}
\,>\,0. 
\label{Grow}
\end{equation}
To simplify the model, 
we disregard the creation of inside excitations, or the out-of-condensate cloud around the Bose-core,
in this article.
Then, \,${\rm const} \simeq  {\rm const}'$.
This means the following terms, 
\begin{equation}
-\,\widetilde{ e_{0}}\,{\rm const}\, N_{\rm o}\,
\bigl(\delta v_{\rm top}\,t\,/\,L_{0}\bigr)^{2}\,+\,\delta e_{0}\,N_{\rm o}\,
\bigl(\delta v_{\rm top}\,t\,/\,L_{0}\bigr)^{2} 
\label{trick}
\end{equation} 
and\, 
\begin{equation}
E_{\rm tail}(t) \simeq \langle \hbar \omega_{\rm x}\rangle \,\delta {N}_{\rm x, \,tail}(t) +
 \langle \hbar \omega_{\rm ph}\rangle \,\delta {N}_{\rm ph, \,tail}(t), 
\label{t_a_i_l}
\end{equation}
have to be included into the (total) energy of the moving packet. 
The time interval $\Delta t$, during which our quasistationary solution for the core\,$+$\,tail can be
used to describe
the transport of the total packet,  can be roughly estimated from the condition \, 
$\vert\delta v_{\rm top}\vert\,\Delta t\,/\,L_{\rm o}\,\simeq \,1$, (see Eq. (\ref{ups1}) and Fig. 3).  
To estimate the excitation energies $\langle \hbar \omega_{\rm x} \rangle $
and $\langle \hbar \omega_{\rm ph}\rangle $ in Eq. (\ref{t_a_i_l}), we have to discuss 
the excitation spectrum of the moving condensate.

\section{Excitations and the tail of soliton}

Within the quasistationary approximation, the following decomposition of
the field operators is used \cite{review},\cite{Griffin}:
$$
\hat{\psi}({\bf x},t) = \Psi_{\rm o}(x,t) 
+ \delta\hat{\psi}({\bf x},t), 
$$
$$
\hat{ u}({\bf x},t) = u_{\rm o}(x,t) +
\delta \hat{u}({\bf x},t),
$$
where the fields $\delta\hat{\psi}$ and $\delta \hat{u}({\bf x},t)$ describe the
out-of-condensate particles.  
To consider the case in which the moving condensate emits excitations during 
the observation time, one has to introduce the fluctuation of the condensate,  
$\delta\Psi_{\rm o}(x,t)$ and $\delta u_{\rm o}(x,t)$, and the fluctuation of
the quasistationary out-of-condensate cloud described by 
$\delta\hat{\psi}'$ and $\delta \hat{u}_{j}'$.
If the number of the ``lost'' particles, $\delta N$, is small 
during the observation time  
(but $\delta N(t)$ is continuously growing), one can consider the moving
coherent packet as a quasistable one. Therefore, we can write the field
decomposition as follows: 
$$
\hat{\psi}({\bf x},t) = \exp\bigl(\,i{\varphi}_{c}(x,t,\psi_{\rm o}(x,t))\,\bigr)\, 
\Bigl\{  \psi_{\rm o}(\,x-v(x,t)t\,) +
\delta\hat{\psi}_{0}(\,x-v(x,t)t,\,{\bf
x}_{\perp},\,t\,)\,\Bigl\}\,+
$$
\begin{equation}
+\,\delta\Psi_{\rm o}(x,t) +
\delta\hat{\psi}'({\bf x},t),
\label{dress}
\end{equation}
$$
\hat{ u}_{j}({\bf x},t) = u_{\rm o}(\,x-v(x,t)t\,)\delta_{1j} +
\delta \hat{u}_{0,\,j}(\,x-v(x,t)t,\, {\bf x}_{\perp},\,t\,)
\,+
$$
\begin{equation}
+\, \delta u_{\rm o}(x,t)\delta_{1j} +
\delta \hat{u}_{j}'({\bf x},t).
\label{dress1}
\end{equation}
In this article, we do not take into account 
the fluctuational parts in this decomposition and, strictly
speaking, we can only estimate the 
the kinetic effects, such as the proposed \,
$\partial_{t}N_{\rm o}(t) \ne 0$ and 
$\partial_{t} \delta n (t) = \partial_{t}\langle \delta\hat{\psi}^{\dag}\,\delta\hat{\psi}\rangle (t) \ne 0$. 
However, the linear quasistationary equations on $\delta\hat{\psi}_{0}$ and 
$\delta \hat{u}_{0,\,j}$ 
can be used to find the excitation spectrum of 
the outside excitations that is required for Eq. (\ref{t_a_i_l}).  
Then, assuming the quasistability of the Bose-core during the observation time $\Delta t$, 
one can switch on the mechanism of occupancy redistribution between the 
macroscopically occupied mode 
$\Psi_{\rm o}(x,t,\,N_{\rm o})$
and the out-of-condensate excitations, e.g., $\delta n_{k}= \langle
\delta\hat{\psi}^{\dag}_{k}\,\delta\hat{\psi}_{k}\rangle $.

Thus, we ``dress'' the quasistationary parts of the out-of-condensate fields by 
Eqs. (\ref{dress}) and (\ref{dress1}) and obtain the following set of equations
($x-v(x,t)t \rightarrow x$):   
$$
i\hbar\,\partial_{t}\,\delta\hat{\psi}_{0}(x,{\bf x}_{\perp},t)\,
=\,\vert {\mu}\vert
\,\delta\hat{\psi}_{0}({\bf x},t)
-\frac{\hbar^{2}}{2m}\,\Delta\delta\hat{\psi}_{0}({\bf x},t)
-\frac{\hbar^{2}}{2m}\,\zeta(v)\,\psi_{\rm o}^{2}(x)\,
\Delta\delta\hat{\psi}_{0}({\bf x},t)\,+
$$
$$
+\,
\left\{ 
(\nu_{0} +  \widetilde{\nu_{0}} + \vert {\mu}\vert \,\zeta(v)\,)
\,\psi_{\rm o}^{2}(x) 
+(2\nu_{1} +  \widetilde{\nu_{1}})\,
\psi_{\rm o}^{4}(x)\,\right\}
\delta\hat{\psi}_{0}(x,{\bf x}_{\perp},t)\,+ 
$$
\begin{equation}
+\, 
\left\{
\nu_{0}\,\psi_{\rm o}^{2}(x) 
+ 2\nu_{1}\,\psi_{\rm o}^{4}(x)\,
\right\}
\delta\hat{\psi}^{\dag}_{0}({\bf x},t)
\,+ 
\,\sigma_{\rm eff}\,
\psi_{\rm o}(x)\,\nabla \delta\hat{\bf u}_{0}(x,{\bf x}_{\perp},t),
\label{delta_psi_}
\end{equation}
$$
\left(\bigl\{\partial_{t}-v(x,t)\,\partial_{x}\bigr\}^{2} -c_{l}^{2}\Delta 
\right)
\delta\hat{ u}_{0,\,x}(x,{\bf x}_{\perp},t)\,-
$$
$$
- \, c_{l}^{2} 2\kappa_{3}\,\partial_{x} u_{\rm o}(x)\, 
\partial_{x}^{2}\delta\hat{ u}_{0,\,x}(x,{\bf x}_{\perp},t)
-  c_{l}^{2} 2\kappa_{3}\,\partial_{x}^{2} u_{\rm o}(x)\,
\partial_{x} \delta\hat{ u}_{0,\,x}(x,{\bf x}_{\perp},t)=
$$
\begin{equation}
=\rho^{-1}\,\tilde{\sigma}_{\rm eff}\,\partial_{x}\!\left( 
\psi_{\rm o}(x)\,\delta\hat{\psi}_{0}(x,{\bf x}_{\perp},t) 
+{\rm h.c.}\,\right).
\label{delta_u1_}
\end{equation}
Here, we simplified the boson-phonon coupling terms and, roughly, 
$\sigma_{\rm eff} \approx \tilde{\sigma}_{\rm eff} \approx \sigma_{0}$.
Note that the interaction between $\delta\hat{\psi}_{0}$ and 
$\delta\hat{ u}_{0,\,x}$ is mediated through the condensate 
${\psi}_{\rm o}(x,t)\cdot \partial_{x} u_{\rm o}(x,t)$
in the linear theory at $T\rightarrow 0$.

For the {\it outside} excitations,
the first approximation of Eqs. (\ref{delta_psi_}) and (\ref{delta_u1_})
consists of taking uncoupled excitons and phonons.
However,  
they live in a half of the medium ($x<0$) and form a 
tail behind the condensate. 
Then, we have the excitonic excitations with 
$$
\hbar \omega_{\rm x}(k_{x})
=\vert \mu \vert + \vert \mu\vert\,(k_{x}L_{0})^{2}, 
\,\,\,\, \vert k_{x}\vert L_{0}<1,
$$
$$
\delta\hat{\psi}(x,t) \sim \bigl(\sqrt{V/2}\,\bigr)^{-1}\, 
\exp(\,i\varphi_{k} + i k_{x}x
-i\omega_{\rm x}t\,)\,\hat{a}_{k},
\,\,\,\,x<0.
$$
This is exactly   
a free exciton with the energy 
$\tilde{E}_{g} + \hbar^{2}(\,k_{0}-\vert k_{x}\vert\,)^{2}/2m $ in the laboratory
frame. It moves behind the condensate with \,
$p_{x}=\hbar \,(k_{0}-\vert k_{x}\vert )>0$. 
For the acoustic phonons, we have ($k'_{x}>0$):  
$$
\hbar \omega_{\rm ph}(k'_{x})=
\hbar\omega_{\rm ac}(k'_{x}) - v\,\hbar k'_{x},
$$
$$
\delta\hat{ u}_{x}(x,{\bf x}_{\perp},t) \sim {\rm const}(k')\,
\exp\bigl(i\varphi_{k'} + ik'_{x}x -i
\{ \omega_{\rm ac}(k'_{x}) - k'_{x}v\}t\, \bigr)\,\hat{b}_{k'} \,+ \,{\rm h.c.},
$$
where \,${\rm const}(k') \propto \bigl(\sqrt{\rho\,(V/2)\,\omega_{\rm ac}(k') }\,
\bigr)^{-1}$.
(After $x\rightarrow x-vt$, we have free phonons in the tail, 
and $k'_{x}$ is the same in the laboratory frame.)
The `resonance' equation for the excitations written 
in the comoving frame has the following form: 
\begin{equation}
\vert \mu \vert + \vert \mu \vert \,(k_{x}L_{0})^{2} = 
\hbar \,(c_{l} \mp v)\,k'_{x},
\end{equation}
and it can be solved easily. Here, we use the following estimate 
for the `relevant' phonons:
$$
\bigl( k'_{x}\,L_{0} \bigr) \approx 0.5\, 
\frac{ \sqrt{ \vert \mu\vert\, \Big/\, mc_{l}^{2}/2 }}{(1\,\mp\,v/c_{l})}.
$$
For $k'_{x}>0$,  we have 
$
\bigl(  k'_{x}\,L_{0} \bigr) \simeq  (1-10)\, 
 \sqrt{ \vert \mu\vert\, \Big/\, mc_{l}^{2}/2 } 
$.
In this article, we do not go into detail 
of how the coupling between the outside
excitons and phonons is formed within  
Eqs. (\ref{delta_psi_}), (\ref{delta_u1_}). 
Our aim is to show that
it is possible to balance the energy of the total packet
by taking into account the growing out-of-condensate tail (\ref{Grow})   
behind the coherent state (\ref{A_Z1}), (\ref{A_Z2}) with leakage (\ref{Leak}).   

Returning to Eq. (\ref{t_a_i_l}), 
we estimate the characteristic energies of excitations as follows:\,
\begin{equation}
\langle \hbar \omega_{\rm x}\rangle \sim \hbar^{2} \tilde{k}_{0}^{2}/2m 
\,\,\,
{\rm and} 
\,\,\,
\langle \hbar \omega_{\rm ph}\rangle \sim {\rm const}_{k}\,\vert \mu\vert, 
\,\,\,{\rm const}_{k} \simeq 2-8.
\end{equation}
(The value of $\langle \hbar \omega_{\rm ph}\rangle $ comes 
from the resonance condition at $k_{{\rm ph}, \,x}>0$.) 
We assume that the tail consists of excitons and phonons, and the 
selfconsistency condition between \,$\psi_{\rm o}^{2}(x/L_{0})\propto 
\Phi_{\rm o}^{2}$ and, roughly,  
$\partial_{x} u_{\rm o}(x/L_{0}) \propto 
\Phi_{\rm o}^{2}$ \,leads to the coherent emission 
of the outside excitons and phonons, and 
$\Phi_{\rm o}^{2}\, \rightarrow \,\Phi_{\rm o}^{2}-\delta\Phi_{\rm o}^{2}(t)$.
Then, we have     
$$
\langle \delta \hat{N}_{\rm x, \,tail} \rangle
\simeq \langle \delta \hat{N}_{\rm ph, \,tail} \rangle. 
$$ 
As a result, the energy of the total packet  
$$
E(t) \approx e_{\rm o}\,N_{\rm o}(t) + \delta e_{0}\,N_{\rm o}(t)\,
\bigl(\delta v_{\rm top}(t)\,t\,/\,L_{0}(t)\,\bigr)^{2} - 
\widetilde{ e_{0}}\,{\rm const}\, N_{\rm o}(t)\,\bigl(\delta v_{\rm
top}(t)\,t\,/\,L_{0}(t)\bigr)^{2}
\,+
$$
\begin{equation}
+ \,\langle\hbar \omega_{\rm x}\rangle 
\,{\rm const}'\,N_{\rm o}(t)\,\bigl(\delta v_{\rm top}(t)\,t\,/\,L_{0}(t)\bigr)^{2}
+ \langle \hbar \omega_{\rm ph}\rangle 
\,{\rm const}'\,N_{\rm o}(t)\,\bigl(\delta v_{\rm top}(t)\,t\,/\,L_{0}(t)\bigr)^{2}
\label{the_energy}
\end{equation}
is conserved if one takes 
$\widetilde{ e_{0}}\,{\rm const} \simeq \delta e_{0}$ \,and \,
${\rm const} \simeq {\rm const}' \simeq 0.1$ in Eqs.
(\ref{trick}), (\ref{t_a_i_l}), and (\ref{the_energy})
(the last value is probably overestimated).

To address the problem of the dynamic stability, one has to
estimate how many particles and phonons are emitted out from 
the moving coherent packet. 
We start from the same initial conditions as in Fig. 2. However, 
the occupations of the tail states start growing immediately as $t>0$ while  
the occupancy of the coherent state starts decreasing. 
Using the conditions 
$E(t)=E_{\rm o}(t=0)$ \,and\, 
$N_{\rm o}(t) + \delta N_{\rm tail}(t) = N_{\rm o}(t=0)$, 
we obtain the dynamics presented on Fig. 3 by 
combination of numerics and analytic estimates.
For example,  
after the  transport has been observed 
during the time interval of $\Delta t$ corresponding to 
$\delta v_{\rm top}\,\Delta {t}\,/\,L_{0} \simeq
(1/4\,-\,1/2)$, the soliton develops the leakage of  
\,$\delta N \sim (1 - 3)\,10^{-2}\,N_{\rm o}$ into the tail excitations, 
see Fig. 3. 
However, the value of  
$\delta v_{\rm top}(N_{\rm o})\,\Delta t\,/\,L_{0}(N_{\rm o})$ starts to decrease 
because of this leakage too. 
For instance, when the adjusted value of  
$\delta v_{\rm top}\bigl(\,N_{\rm o}(t)\,\bigr)\,
\Delta t'\,/\,L_{0}\bigl(\,N_{\rm o}(t)\,\bigr)
\simeq 1$, the leakage is esimated as \,
$\delta N \simeq (1-2)\,10^{-1}\,N_{\rm o}$, see Fig. 3.   
Therefore,  the condensate can be considered 
as a quasistable one during the time interval of \,$\simeq \Delta t'$.
The following formulas can be used for estimates
\begin{equation}
\Phi_{\rm o}^{2}(t)\,S_{\perp}\,2L_{0}(t) \simeq N_{\rm o}(t),
\end{equation}
\begin{equation}
\delta n_{\rm tail}(t)\,S_{\perp}\,\tilde{L} \simeq  q(t)\,N_{\rm o},
\end{equation}
where
$\delta n_{\rm tail}(t)\simeq q'\,\Phi_{\rm o}^{2}(t)$, \,$\tilde{L} = \tilde{q}\,L_{0}$.
If $q(t) \ll 1$ (e.g., $q \le 10^{-1}$ within the time 
interval $\Delta t'$ on Fig. 3) 
and 
$\tilde{q}\gg 1$ (e.g., $\tilde{q} \sim 10-50$), 
one can roughly estimate the dimensionless 
\,$q'$ \,as\,  $\sim 10^{-1}\,q$.
The dynamics presented on Fig. 3 is in a qualitative agreement with 
experimental data \cite{cuprous}.
(It is interesting to mention that the similar problems of 
dynamic stability appear in the theory of the so-called 
Embedded  Solitons in nonlinear optics  
 \cite{Yang}.)

%
\begin{figure}[p]
\begin{center}
\leavevmode
\epsfxsize = 350pt
\epsfysize = 200pt
\epsfbox{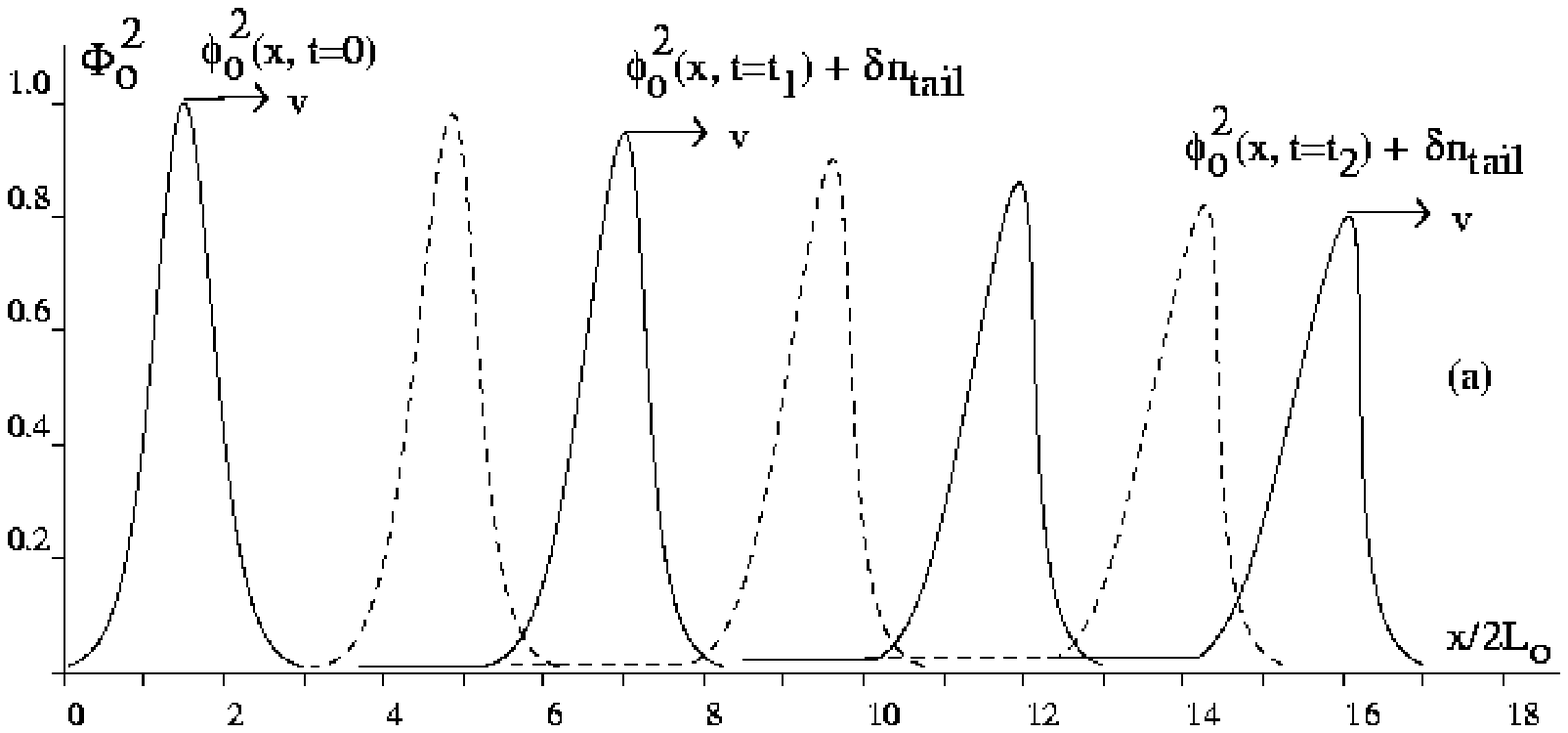}
\end{center}
\end{figure}
\begin{figure}[p]
\begin{center}
\leavevmode
\epsfxsize = 350pt
\epsfysize = 125pt
\epsfbox{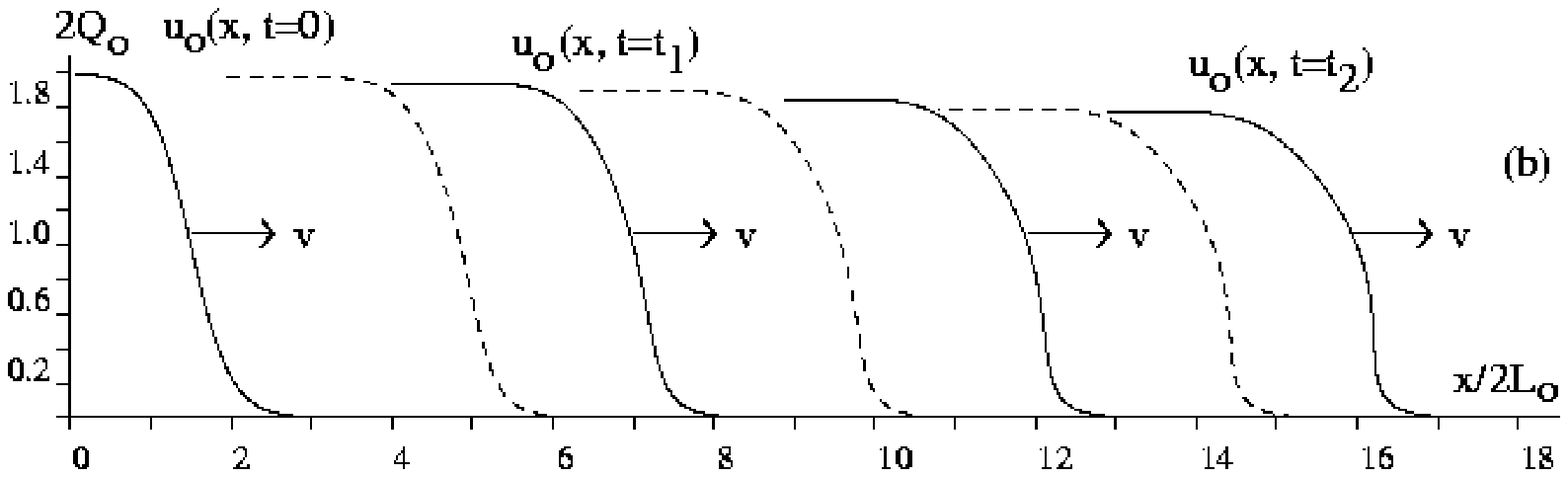}
\end{center}
\caption{
To model transport properties of the boson-phonon soliton 
in the case of effective dissipation (the ``leakage''),
we started from the symmetric soliton without a tail as an 
initial condition at $t=0$.
Dynamics of the boson (exciton) part of the packet is presented on Fig. 3\,(a) 
in the form of moving \,
$\vert \Psi_{\rm o}(\,x-v(x,t)t\,)\vert ^{2} + \delta n(x,t)$, where \,
$\delta n(x,t)\simeq \delta n_{\rm tail}$ at $\bar x < -(2-3)\,L_{0}$. 
The coherent phonon part (a moving kink of the displacement field) is depicted on 
Fig. 3\,(b); the phonon part of the tail 
$\langle (\partial_{x} \delta \hat{u}_{x})^{2} \rangle (t) \ne 0$ 
is not presented on this figure.  
The initial value of the interaction 
parameter $\zeta(v)\,\Phi_{\rm o}^{2} \simeq \delta v_{\rm top}/v$ 
is taken to be 
$+0.05$. 
Then, the visible changements 
occur after the packet has traveled the distance of \,$(20-30)\, L_{0}$, which corresponds
to the effective value of $ \delta v_{\rm top}(t)\,\Delta t / L_{0}(t) \sim 1$.  
Note that the energy of the  total moving packet 
(coherent part\,$+$\,tail) is conserved 
at $T\rightarrow 0$.  
}
\label{soliton_tail}
\end{figure}

\newpage

\section{Conclusion}

When a quasistationary exciton-phonon soliton moves in a periodic medium, 
an asymmetric form of the ballistic signal can be developed as a dynamic effect
due to exciton-phonon interaction.
This effect has the tendency to be accumulated with time,
and it  can be clearly seen after 
the packet has traveled the distance of $n\,L_{0}$, \,$n\gg 1$.
Here, the width of the packet is $\simeq (3-6)\,L_{0}$, and  
the value of $n$ depends on the strength of relevant interaction
parameters, such as \,$\vartheta_{0}$ and $\sigma_{0}$. It can be 
estimated by the following formula 
(see Eq. (\ref{howMANY})\,):
$$ n \simeq \bigl(\,
\vert \delta v_{\rm top}\vert \Big/ v \,\bigr)^{-1} \sim 10^{3}-10^{2}.
$$

To a first approximation, we introduced the coherent part of the packet
and described its dynamics  
by use of the selfconsistent generalization of \,$v={\rm
const}\,\rightarrow\, v(x,t)$ \,and\, 
$\varphi_{\rm c}(x,t)=k_{0}x-\omega_{0}t\,\rightarrow \,
\tilde{\varphi}_{\rm c}(x,t)$ in such a way: 
$$
v \,\stackrel{\vartheta_{0}\ne 0}{\longrightarrow}\, 
v(\,\partial_{x}u_{\rm o}(x,t)\,) \longrightarrow 
v(\,\vert{\Psi}_{\rm o}(x,t)\vert^{2}\,)
$$ 
$$
\varphi_{\rm c}(x,t) \,\stackrel{\vartheta_{0}\ne 0}{\longrightarrow}\, 
\tilde{\varphi}_{\rm c}(x,t,\,
\,\partial_{x}u_{\rm o}(x,t)\,)
\longrightarrow
\varphi_{\rm c}(x,t,\,
\vert{\Psi}_{\rm o}(x,t)\vert^{2}\,).
$$
This substitute was applied within the quasistationary approximation 
to simplify and solve the corresponding dynamic equations on 
the coherent Bose-core of the total packet.
The incoherent part of it turned out to be equally important
in dynamics of the total packet. Indeed, the quasistationary 
solution for the Bose-core 
cannot move in a periodic medium as a single soliton, 
but with the leakage into the out-of-condensate excitation states. 
Thus, the question on whether a moving conserving  
solution (i.e., a soliton conserving at least 
$E_{\rm o}$, $P_{{\rm o},\,x}$, and $N_{\rm o}$) 
exists in the model with \,$\vartheta_{0}\ne 0$ and $\sigma_{0}\ne 0$ 
remains open. 
In this article, we showed that an exciton-phonon ``comet'' 
is a better image for the moving ballistic packet, 
and the core of it (a kind of the ``nucleus'' of
such a  comet) 
can be modeled by a coherent state, or a condensate. 

The ansatz we used for the soliton-like solution can be applied 
within the validity of
the adiabatic hypothesis (i.e., it is 
not the exact one of the dynamic equations). 
As a result, we could obtain an explicit dependence on time 
(\,$\propto \bigl(\delta v_{\rm top}\,t\,/\,L_{0}\bigr)^{2}$\,) for 
the exact energy 
and momentum of the moving soliton if it were the only one component 
of the moving packet.
However, 
the 
technique of Bose-Einstein condensation allows to proceed with the approximate 
solution obtained in this article.  
It is known that the interaction between Bose-condensate and non-condensed 
particles leads to an effective dissipation for the condensate wave function at 
$T\ne 0$  
\cite{Stoof}.
We assume that  
an effective dissipation appears naturally in description of the  transport 
properties  
of the coherent part,
$\Psi_{\rm o}({ x},t)$ and $\partial_{x} u_{\rm o}({ x},t)$, 
at $T\rightarrow 0$. 
If the initial state of the packet was taken as a  
pure coherent state with a nonzero momentum,
the moving condensate emits 
excitations, i.e., 
$N_{\rm o}\,\rightarrow\,  N_{\rm o} - \delta N(t)$ during the observation time.
Then, the out-of-condensate cloud of collective excitations and, in particular,  
the tail consisting of the out-of-condensate excitons and phonons  
grow around and behind the moving coherent state, e.g.,  
$\langle \,\delta \hat{n}(x,t)\,\rangle \ne 0$, \,$x < - 3\,L_{0}$.

We argue that
the emission effect can make the dynamics of the total moving packet
conservative. Therefore, the coherent part  of 
such a packet has to be considered as a 
quasistable core (a ``nucleus'' of the ``comet'') of the moving
exciton-phonon droplet 
provided \,$\delta N_{\rm tail }(t) \ll  N_{\rm o}(t)$ 
during the observation time.
In addition, the coherent phase $\varphi_{\rm c}(x,t)$ prescribed to the 
macroscopic wave 
function $\Psi_{\rm o}(x,t)$ cannot be taken as a regular field 
under such conditions because of dephasing effects \cite{Leggett}.
Its fluctuations  has to be taken into account to clarify 
the coherent properties of the Bose-core.   
On the other hand,  
the rigorous approach to the dynamics of solitons with emission lies 
beyond the quasistationary approximation we used in this articles. 
A set of kinetic equations has to be applied to treat this problem in 
detail.

\section{Acknowledgements}

One of the authors (D.R.) thanks I. Loutsenko for critical reading of the manuscript.


\begin{thebibliography}{100}


\bibitem{Dodd}
R. K. Dodd, J. C. Eilbeck, J. D. Gibbon, H. C. Morris, 
{\it Solitons and Nonlinear Wave Equations}, (Academic Press, 1982).

\bibitem{Infeld}
E. Infeld, G. Rowlands, {\it Nonlinear Waves, Solitons, and Chaos},
(Cambridge University Press,
Cambridge, 1990).

\bibitem{Bishop}
A. R. Bishop, M. G. Forest, D. W. Mclaughlin, and E. A. Overman II, 
Physica {D} {\bf 23}, 293 (1986).

\bibitem{Enns}
D. E. Edmundson and R. H. Enns, Phys. Rev. A {\bf 151}, 2491 (1995).

\bibitem{Pelin}
Y. S. Kivshar, D. E. Pelinovsky, Phys. Rep. {\bf 331}, 200 (2000).

\bibitem{numeric}
B. Fornberg, G. B. Whitham, Phil. Trans. Royal. Soc. London {\bf 289}, 373 (1978);
\newline
S. B. Wineberg, J. McGrath, E. Gabl, L. R. Scott, C. Southwell, Comp. Phys. {\bf 97}, 311 (1991);
\newline
C. Cercignani, D. H. Sattinger, {\it Scaling Limits and Models in Physical Processes},
(Birkh\"auser, Boston-Basel-Berlin, 1998).

\bibitem{review}
{\it Bose-Einstein
Condensation}, edited by
A. Griffin, D. W. Snoke and S. Stringari (Cambridge University Press,
Cambridge, 1995);
\newline
F. Dalfovo, S. Giorgini, L. P. Pitaevskii, and S. Stringari 
Rev. Mod. Phys. {\bf 71}, 463 (1999).

\bibitem{Newell}
A. Hasegawa, {\it Optical Solitons in Fibers}, (Springer-Verlag, Berlin, 1989);
\newline
A. C. Newell, J.V. Moloney. {\it Nonlinear Optics}, (Addison-Wesley, Redwood City, 1992).

\bibitem{Benjamin}
T. B. Benjamin, J. Bona, J.J. Mahoney, Phil. Trans. Royal. Soc. London A. {\bf 272}, 47 (1972).

\bibitem{Braun}
O. M. Braun, Y. S. Kivshar, Phys. Rep. {\bf 306}, 1 (1998).

\bibitem{Davydov}
A. S. Davydov, {\it Solitons in Molecular Systems}, (Reidel, Dordrecht, 1984);
\newline
{\it Davydov's Soliton Revisited}, edited by
P. L. Christiansen, A. C. Scott, NATO ASI Series B: Physics {\bf 243},
(Plenum Press, 1990).

\bibitem{Spin}
A. J. Heeger, S. Kivelson, J. R. Schrieffer, and W. P. Su, 
Rev. Mod. Phys {\bf 60}, 781 (1988);
\newline
D. Augier, D. Poilblanc, E. Sorensen, I. Affleck, Phys. Rev. B {\bf 58}, 9110 (1998).

\bibitem{CDW} 
{\it Charge Density Waves in Solids}, L. Gor'kov and G. Gr\"uner eds., 
(Elsevier Sci. Publ., Amsterdam, 1989).

\bibitem{excitons}
J. L. Lin, J. P. Wolfe, Phys. Rev. Lett. {\bf 71}, 122
(1993);
\newline
E. Fortin, S. Fafard, A. Mysyrowicz, Phys. Rev. Lett.
{\bf 70}, 3951 (1993);
\newline
H. Kondo, H. Mino, I. Akai, and T. Karasawa, Phys. Rev. B {\bf 58}, 13835 (1998);
\newline
L. V. Butov, A. I. Filin, Phys. Rev. B {\bf 58}, 1980 (1998);
\newline
V. Negoita, D. W. Snoke, K. Eberl, Phys. Rev. B {\bf 60}, 2661 (1999).

\bibitem{cuprous}
A. Mysyrowicz, E. Benson, and E. Fortin, 
Phys. Rev. Lett.  {\bf 77}, 896 (1996);
\newline
E. Benson, E. Fortin, A. Mysyrowicz, Sol. Stat. Comm. {\bf 101}, 313
(1997);
\newline
E. Benson, E. Fortin, B. Prade and A. Mysyrowicz, 
Europhys. Lett. {\bf 40}, 311 (1997);
\newline
E. Fortin, E. Benson, and A. Mysyrowicz,  Electrochem. Soc. Proceedings {\bf 98-25}, 1 (1998).

\bibitem{in_action}
S. F. Mingaleev, P. L. Christiansen, Y. B. Gaididei, M. Johansson, K. O. Rasmussen,
J. of Biolog. Phys. {\bf 25}, 41 (1999);
\newline
A. V. Zolotaryuk, K. H. Spatschek, A. V. Savin, Phys. Rev. B {\bf 54}, 266
(1996); 
\newline
P. L. Christiansen, Y. B. Gaididei, S. F. Mingaleev, cond-mat/0003146.

\bibitem{Tusin}
E. A. Bartnik, J. A. Tuszy\'nski, Phys. Rev. E {\bf 48}, 1516 (1993);

\bibitem{CMP}
{\it Microscopic Aspects of Nonlinearity in Condensed Matter}, ed. by A. R. Bishop, V. L.
Pokrovsky, and A. Tognetti, (Plenum, New York, 1991).

\bibitem{Loutsenko}
I. Loutsenko, D. Roubtsov, Phys. Rev. Lett. {\bf 78}, 3011 (1997);\,
{\it ibid}. {\bf 84}, 3503 (2000).

\bibitem{Brasil}
A. R. Vasconcellos, M. V. Mesquita, and R. Luzzi, Europhys. Lett. {\bf 49}, 637 (2000).
 
\bibitem{Zakharov} 
V. E. Zakharov, Sov. Phys. JETP {\bf 35}, 908 (1972).

\bibitem{Sulem}
C. Sulem, P.-L. Sulem, {\it The nonlinear Schr\"odinger equation. Self-Focusing and Wave Collapse.},
(Springer-Verlag, New York Inc., 1999).

\bibitem{Berge}
L. Berg\'e, Phys. Rep. {\bf 303}, 259 (1998).

\bibitem{NMPZ}
S. Novikov, S. V. Manakov, L. P. Pitaevskii, V. E. Zakharov, {\it Theory of Solitons. The Inverse Scattering
Method}, (Consultants Bureau, New York and London, 1984).

\bibitem{Ablowitz}
M. J. Ablowitz, P. A. Clarkson, {\it Solitons, Nonlinear Evolution Equations and Inverse Scattering}, 
(Cambridge University Press, New York, 1991).

\bibitem{Melnik}
V. K. Mel'nikov, Comm. Math. Phys. {\bf 112}, 639 (1987);
J. Math. Phys. {\bf 28}, 2603 (1987).

\bibitem{tail_exp}
G. B. Whitham, 
{\it Linear and Nonlinear waves}, (John Willey$\&$Sons, New York, 1973).

\bibitem{bio}
S. Georghiou, T. D. Bradrick, A. Philippetis and J. M. Beechem, Biophysical J. {\bf 70}, 1909
(1996);
\newline
S. O. Kelley, J. K. Barton, Science {\bf 283}, 375 (1999);
\newline 
J. Schiller, G. Major, H. J. Koester, Y. Schiller,  Nature {\bf 404}, 285 (2000).
\newline
A. Xie, L. van der Meer, W. Hoff, R. H. Austin,  Phys. Rev. Lett. {\bf 84}, 5435 (2000).

\bibitem{Agrawal}
G. P. Agrawal, {\it Nonlinear Fiber Optics}, (Second Edition, Academic Press,
New York 1995), and references therein.

\bibitem{Kaup}
D. J. Kaup, T. I. Lakoba, B. A. Malomed, J. Opt. Soc. Am. B {\bf 14}, 1199
(1997).

\bibitem{Camassa}
R. Camassa, D. D. Holm, Phys. Rev. Lett. {\bf 71}, 1661 (1993);
\newline
P. Rosenau, Phys. Rev. Lett. {\bf 73}, 1737 (1994).

\bibitem{Boyd}
J. P. Boyd, {\it Weakly Nonlocal Solitary Waves and Beyond -- All-Orders
Asymptotics} (Kluwer, Dodrecht, Boston, London, 1998).



\bibitem{Inoue}
J. Inoue, T. Brandes, and A. Shimizu, Phys. Rev. B {\bf 61}, 2863 (2000).

\bibitem{exc_transp} 
C. Rocke, S. Zimmermann, A. Wixforth, J. P. Kotthaus, Phys. Rev. Lett. {\bf 78}, 4099 (1997).

\bibitem{ScLength}
J. Shumway and D. M. Ceperley, cond-mat/9907309.

\bibitem{Hennig}
D. Hennig, G. P. Tsironis, Phys. Rep. {\bf 307}, 333 (1999).

\bibitem{RL}
D. Roubtsov, Y. L\'epine, Phys. Stat. Sol. B {\bf 210}, 127 (1998);
\,Phys. Rev. B {\bf 61}, 5237 (2000).

\bibitem{thermo}
A. L. Ivanov, C. Ell, and H. Haug, Phys. Rev. E {\bf 55}, 6363 (1997);
Phys. Rev. B {\bf 57}, 9663 (1998).

\bibitem{STikho}
A. E. Bulatov, S. G. Tikhodeev, Phys. Rev. B {\bf 46}, 15058 (1992);
\newline
S. G. Tikhodeev, G. A. Kopelevich, N. A. Gippius, 
Phys. Stat. Sol. B {\bf 206}, 45 (1998).

\bibitem{elastic}
A. J. Sievers and S. Takeno, Phys. Rev. Lett. {\bf 61}, 970 (1988);
\newline
S. Flach, C. R. Wills, Phys. Rep. {\bf 295}, 181 (1998); 
\newline
G. Huang and B. Hu,  Phys. Rev. B {\bf  58},  9194 (1998).

\bibitem{Griffin}
A.~Griffin, Phys. Rev. B {\bf 53}, 9341 (1996); 
in {\it Bose-Einstein Condensation in Atomic Gases}, edited by M. Inguscio, S. Stringari
and C. Wieman (Italian Physical Society, 1999).
\newline
E. Zaremba, A. Griffin, T. Nikuni, Phys. Rev. A {\bf 57}, 4695 (1998).

\bibitem{Tilley}
D. R. Tilley, J. Tilley, {\it Superfluidity and Superconductivity}
(A. Hilger, Bristol, 1990).

\bibitem{Salerno}
M. J. Ablowitz, J. F.  Ladik, J. Math Phys. {\bf 17}, 1011 (1976);
\newline
M. Salerno, Phys. Rev. A {\bf 46}, 6856 (1992).

\bibitem{Pomeau}
C. Josserand, Y. Pomeau, S. Rica,  Phys. Rev. Lett. {\bf 75}, 3150 (1995).

\bibitem{cub-quin}
D. Mihalache, D. Mazilu, L.-C. Crasovan. B. A. Malomed, F. Lederer, 
Phys. Rev. E {\bf 61}, 7142 (2000).

\bibitem{Kong}
D. Kong, Phys. Lett. A {\bf 196}, 301 (1995);
\newline
B. Dey, A. Khare, C. N. Kumar, hep-th/9510054.

\bibitem{hydro}
G. Baym, B. Link,  Phys. Rev. Lett. {\bf 69}, 2959 (1992);
\newline
E. Zaremba, T. Nikuni, A. Griffin, J. Low Temp. Phys. {\bf 116}, 277 (1999);


\bibitem{damping} 
L. P. Pitaevskii, S. Stringari, Phys. Lett. A {\bf 235}, 398 (1997);
\newline
T. Nikuni, A. Griffin, cond-mat/0009333; 
\,J. E. Williams, A. Griffin, cond-mat/0003481. 

\bibitem{Yang}
J. Yang, B. A. Malomed, D. J. Kaup, Phys. Rev. Lett. {\bf 83}, 1959 (1999).

\bibitem{Stoof}
H. T. C. Stoof, J. Low Temp. Phys. {\bf 114}, 11 (1999);
\newline
M. J. Bijlsma, H. T. C. Stoof, cond-mat/0007026.

\bibitem{Leggett}
A. J. Leggett, F. Sols, Foundations of Phys. {\bf 21}, 353 (1991); 
Phys. Rev. Lett. {\bf 81}, 1344 (1998).

\end{thebibliography}
\end{document}